\documentclass[12pt]{article}
\usepackage{amsmath,amssymb,amsthm,amsxtra,overpic,bbm,bm,epsfig,ulem,multirow}
\usepackage{color,cite}
\usepackage{epstopdf}
\usepackage{booktabs}
\usepackage{footnote}
\usepackage{rotating}
\usepackage{hyperref}
\usepackage{url}
\usepackage{cite}
\usepackage{tikz}
\usetikzlibrary{arrows,shapes,snakes,automata,backgrounds,petri}
\usetikzlibrary{decorations.pathmorphing}
\usetikzlibrary{decorations.markings}
\usepackage{lineno}

\def\thefootnote{\fnsymbol{footnote}}

\usepackage{array}

\begin{document}

\vspace{0.2cm}

\begin{center}
{\Large\bf Neutral-current background induced by atmospheric neutrinos at large liquid-scintillator detectors:} \\
\vspace{0.0cm}
{\Large\bf \underline{I. model predictions}}
\end{center}

\vspace{0.2cm}

\begin{center}
{\bf Jie Cheng~$^{a}$}~\footnote{Email: chengjie@ihep.ac.cn},
{\bf Yu-Feng Li~$^{a,~b}$}~\footnote{Email: liyufeng@ihep.ac.cn},
{\bf Liang-Jian Wen~$^{a}$}~\footnote{Email: wenlj@ihep.ac.cn},
{\bf Shun Zhou~$^{a,~b}$}~\footnote{Email: zhoush@ihep.ac.cn} \\
\vspace{0.2cm}
{$^a$Institute of High Energy Physics, Chinese Academy of Sciences, Beijing 100049, China}\\
{$^b$School of Physical Sciences, University of Chinese Academy of Sciences, Beijing 100049, China}
\end{center}

\vspace{1.5cm}

\begin{abstract}
The experimental searches for diffuse supernova neutrino background and proton decay in next-generation large liquid-scintillator (LS) detectors are competitive with and complementary to those in the water-Cherenkov detectors. In this paper, we carry out a systematic study of the dominant background induced by atmospheric neutrinos via their neutral-current (NC) interactions with the $^{12}{\rm C}$ nuclei in the LS detectors. The atmospheric neutrino fluxes at the location of Jiangmen Underground Neutrino Observatory (JUNO) are used, as the JUNO detector is obviously a suitable representative for future LS detectors. Then, we implement the sophisticated generators \texttt{GENIE} and \texttt{NuWro} to simulate the neutrino interactions with the carbon nuclei, and the package \texttt{TALYS} to deal with the deexcitations of final-state nuclei. Finally, the event rates for the production of additional nucleons, $\gamma$'s, $\alpha$'s, pions and kaons are obtained and categorized, and the systematic uncertainty of the NC background represented by a variety of data-driven nuclear models is estimated. The implications of the NC background from atmospheric neutrinos for the detection of diffuse supernova neutrino background and proton decay are also discussed.
\end{abstract}


\def\thefootnote{\arabic{footnote}}
\setcounter{footnote}{0}

\newpage

\section{Introduction}\label{sec:intro}

Recent years have seen very impressive progress in neutrino oscillation experiments, which has not only revealed the fundamental properties of neutrino masses and leptonic flavor mixing, but also provided us with more powerful detectors for neutrino astronomy and the experimental searches for other rare events. Among various physics goals of future large-scale neutrino detectors, the discoveries of the long-sought diffuse supernova neutrino background (DSNB) as well as proton decay
feature among the most important.

As about two dozens events of neutrinos from the core-collapse explosion of Supernova (SN) 1987A have been observed in Kamiokande-II~\cite{Hirata:1987hu}, IMB~\cite{Bionta:1987qt} and Baksan~\cite{Alekseev:1988gp} experiments, there exists a guaranteed source of neutrinos from all the SN explosions occurred in the visible Universe over its whole history of evolution, i.e., the DSNB. The integrated flux of diffuse SN neutrinos carries very useful information about the cosmological evolution, the average energy spectrum of core-collapse SN neutrinos, the formation rate of massive stars and the rate of failed SNe~\cite{Ando:2004hc, Beacom:2010kk, Lunardini:2010ab}. Hence the detection of DSNB has been one of the primary goals for the water-Cherenkov (wCh) detector Super-Kamiokande (SK) for a long time~\cite{Malek:2002ns, Bays:2011si, Zhang:2013tua, Zhang:2015zla}. So far, no appreciable inverse-beta-decay (IBD) signal events $\overline{\nu}^{}_e + p \to e^+ + n$ of the DSNB $\overline{\nu}^{}_e$ have been found in the SK-IV phase with the tagging of neutrons via their captures on hydrogen, leading to an upper bound on the differential flux $\phi^{}_{\overline{\nu}^{}_e} (E^{}_{\overline{\nu}_e}) < 5~{\rm cm}^{-2}~{\rm s}^{-1}~{\rm MeV}^{-1}$ at $E^{}_{\overline{\nu}_e} = 19~{\rm MeV}$~\cite{Zhang:2013tua}. In the near future, the Gadolinium-doped {phase SK-VI} will feature greatly improved efficiency for neutron tagging and hence significantly reduce background levels, {increasing sensitivity of the DSNB}~\cite{Beacom:2003nk, Watanabe:2008ru, Horiuchi:2008jz, Labarga:2018fgu}.  One of the dominant backgrounds for the DSNB searches at SK comes from long-lived isotopes induced by the cosmic-ray muon spallation, which has been systematically studied in a series of papers by Li and Beacom~\cite{Li:2014sea, Li:2015kpa, Li:2015lxa}.

Compared to the wCh detectors, the liquid-scintillator (LS) detectors offer lower energy thresholds and higher energy resolutions. The DSNB search in {LS detectors} has been previously taken up by the KamLAND collaboration~\cite{Collaboration:2011jza}. The observation of extraterrestrial $\overline{\nu}^{}_e$ at KamLAND is { highly} consistent with the expected background, which is dominated by the neutral-current (NC) interactions of atmospheric neutrinos with the carbon nuclei, setting an upper limit of $139~{\rm cm}^{-2}~{\rm s}^{-1}$ on the total flux of DSNB $\overline{\nu}^{}_e$ in the analyzed energy range $8.3~{\rm MeV} < E^{}_{\overline{\nu}^{}_e} < 31.8~{\rm MeV}$~\cite{Collaboration:2011jza}. For future large LS detectors, such as JUNO~\cite{An:2015jdp} and LENA~\cite{Wurm:2011zn}, the NC background stemming from atmospheric neutrinos demands a dedicated investigation, which is critically important for the DSNB discovery~\cite{Mollenberg:2014pwa, Wei:2016vjd, Priya:2017bmm, Moller:2018kpn}. Some comments on the signals and backgrounds for the DSNB searches in the LS detectors are helpful.
\begin{itemize}
\item Due to its large cross section, the IBD is the ideal channel for the detection of the DSNB $\overline{\nu}^{}_e$ component. Moreover, the time coincidence of the prompt and delayed signals arising respectively from the final-state $e^+$ and $n$ perfectly eliminates the single-event backgrounds, such as the single signals from the radioactivity of detector materials and the recoiled electrons from solar neutrino interactions.

\item In the low-energy part of the DSNB $\overline{\nu}^{}_e$ spectrum, an important irreducible background originates from those $\overline{\nu}^{}_e$'s emitted from nearby nuclear reactors. However, the flux of reactor neutrinos is highly suppressed above the neutrino energy of around 10 MeV.
    The high-energy part of the background is mainly composed of the IBD interactions of the atmospheric $\overline{\nu}^{}_e$ as well as the charged-current (CC) interactions of atmospheric neutrinos with ${^{12}{\rm C}}$ nuclei in LS, where copious neutrons, protons, $\gamma$'s and $\alpha$'s are generated and can contaminate the IBD signals. The CC background induced by atmospheric neutrinos can be essentially removed by shrinking the {signal} energy window. Therefore, the energy range of interest is usually restricted to the region between two intrinsic $\overline{\nu}^{}_e$ backgrounds from reactor and atmospheric neutrinos. { It is worthwhile to mention that the DSNB sensitivity is very sensitive to the lower bound of the energy window, but the upper bound is not terribly important.} The exact window depends to a large extent on the chosen detector and the control of various backgrounds.

\item Unlike the wCh detectors~\cite{Li:2014sea, Li:2015kpa, Li:2015lxa}, the cosmic-ray muon spallation related backgrounds are well under control in the LS detectors by implementing the muon veto. Fast neutron background generated by muon spallation outside the detector can be removed by cutting the outer layer of the detector, implying a slight reduction of the fiducial volume. Thus a balance between the signal loss and the reduction of backgrounds should be made.

\item In the chosen energy window, we are finally left with the dominant NC background caused by atmospheric neutrinos. Different from the CC interactions, where the associated final-state charged leptons carry away most of the initial-state neutrino energies and deposit their energies in the LS detectors, the final-state neutrinos in the NC interactions escape from detection. Meanwhile, the produced neutrons, protons, $\alpha$'s and residual light nuclei, whose deexcitations result in high-energy gamma rays, contribute to the main background.
\end{itemize}

Different from the DSNB, that is undoubtedly presented but yet to be discovered, {proton decay is predicted by the grand unified theories (GUTs)~\cite{Georgi:1974sy, Fritzsch:1974nn} of particle physics, and such an observation would offer exciting evidence for new fundamental theories.} Historically, the experimental searches for proton decay $p \to e^+ + \pi^0$ in the wCh detectors~\cite{Arisaka:1985ce,Bionta:1983kz} actually accelerated the development of neutrino physics. The latest limits for proton decay at the SK can be found in Refs.~\cite{Abe:2014mwa, Miura:2016krn, Tanaka:2020emn}, while the studies of $p \to K^+ + \overline{\nu}$ in the LS detectors have been carried out in Refs.~\cite{Undagoitia:2005uu, TheKamLAND-Zen:2015eva}. Compared to those of the DSNB, the experimental signals of proton decay and the relevant backgrounds occur at higher energies. This {necessitates the study of NC background} over a broad energy range.

Motivated by the prominent importance of the experimental searches for DSNB and proton decay, in the present work we perform a systematic study of the NC background induced by atmospheric neutrinos. For this purpose, several ingredients are required. {First comes the precise calculation of the fluxes of atmospheric neutrinos, which originate from the interactions of high-energy cosmic rays with the nuclei in the Earth's atmosphere~\cite{Gaisser:2002jj}, and depend on the specific detector site under consideration.}
To be concrete, we choose the JUNO site, but the details of the JUNO detector are not required. Secondly, the interactions between atmospheric neutrinos of energies ranging from MeV to 10 GeV and the target $^{12}{\rm C}$ nuclei in the LS detectors are to be modeled. We employ the widely-used generators \texttt{GENIE}~\cite{Andreopoulos:2009rq} and \texttt{NuWro}~\cite{Golan:2012rfa} for the neutrino interactions, and the package \texttt{TALYS}~\cite{Koning:2005ezu} for the deexcitations of the final-state nuclei.
{By using a variety of data-driven nuclear models, we are able to obtain reliable predictions of the NC background in the searches of DSNB and proton decay, and estimate the associated systematic uncertainties.}
Finally we would like to mention that NC interactions investigated here are also a significant background for searches of neutrinos from the dark matter annihilation~\cite{PalomaresRuiz:2007eu,Guo:2015hsy}.

The remaining part of this paper is organized as follows. In Sec.~\ref{sec:framework}, we outline our strategy for the practical calculations of the NC background in the LS detectors, where all the necessary ingredients are offered and explained. Then, the final results of the NC background are given in Sec.~\ref{sec:results}, and the resultant energy spectra of protons, neutrons, $\gamma$'s, $\alpha$'s and others are also provided, which may be useful for other general-purpose studies. In addition, the NC background rates are estimated for the searches for DSNB and proton decay. Finally, we summarize our main results and conclude in Sec.~\ref{sec:summary}.

\section{Strategy for Calculations}\label{sec:framework}

\subsection{Atmospheric Neutrino Fluxes}\label{subsec:flux}

In the first place, the fluxes of atmospheric neutrinos $\nu^{}_\mu$, $\overline{\nu}^{}_\mu$, $\nu^{}_e$ and $\overline{\nu}^{}_e$ must be calculated as precisely as possible. Since we are concerned about NC interactions of atmospheric neutrinos with the carbon nuclei in the LS detectors, which are insensitive to neutrino flavors, the flavor conversions of atmospheric neutrinos will have no impact on the final results. In our calculations, these fluxes have been taken from the latest results by the Honda group~\cite{Honda:2015fha}, {and the associated uncertainties can be reduced by using the accurately measured atmospheric muon flux~\cite{Honda:2006qj,Honda:2019ymh}. A brief summary of the flux calculations in Refs.~\cite{Honda:2006qj,Honda:2015fha, Honda:2019ymh} follows.}

The calculation of atmospheric neutrino fluxes is carried out by simulating realistic interactions of primary cosmic rays with the atoms in the Earth's atmosphere, the propagation of primary and secondary particles in the geomagnetic fields, {and} the decays and interactions of produced mesons and subsequent muons. First, the temperature and air density profiles of the Earth's atmosphere are provided by the NRLMSISE-00 model~\cite{Picone}, which improves the density profile in U.S.-standard 1976~\cite{US} by taking account of the time variation and the position dependence around the Earth. Second, the model of primary cosmic rays has been constructed by incorporating recent precision measurements by AMS02~\cite{Aguilar:2015ooa, Aguilar:2015ctt} and other experiments~\cite{Adriani:2013as, Abe:2017yrg}. Third, the hadronic interaction models~\cite{Honda:2006qj, Niita:2006zz} are implemented to deal with the collisions between the primary cosmic rays and the atoms in the atmosphere, and then the realistic \texttt{IGRF} geomagnetic model~\cite{igrf} is utilized to perform three-dimensional simulations of the propagation of cosmic rays and their secondaries~\cite{Honda:2001xy}. Finally, the neutrino fluxes are obtained by following their motion in the atmosphere and selecting only those registered in the virtual detectors at the chosen experimental site~\cite{Honda:1995hz}.
\begin{figure}[!t]
\begin{center}
\includegraphics[width=0.7\textwidth]{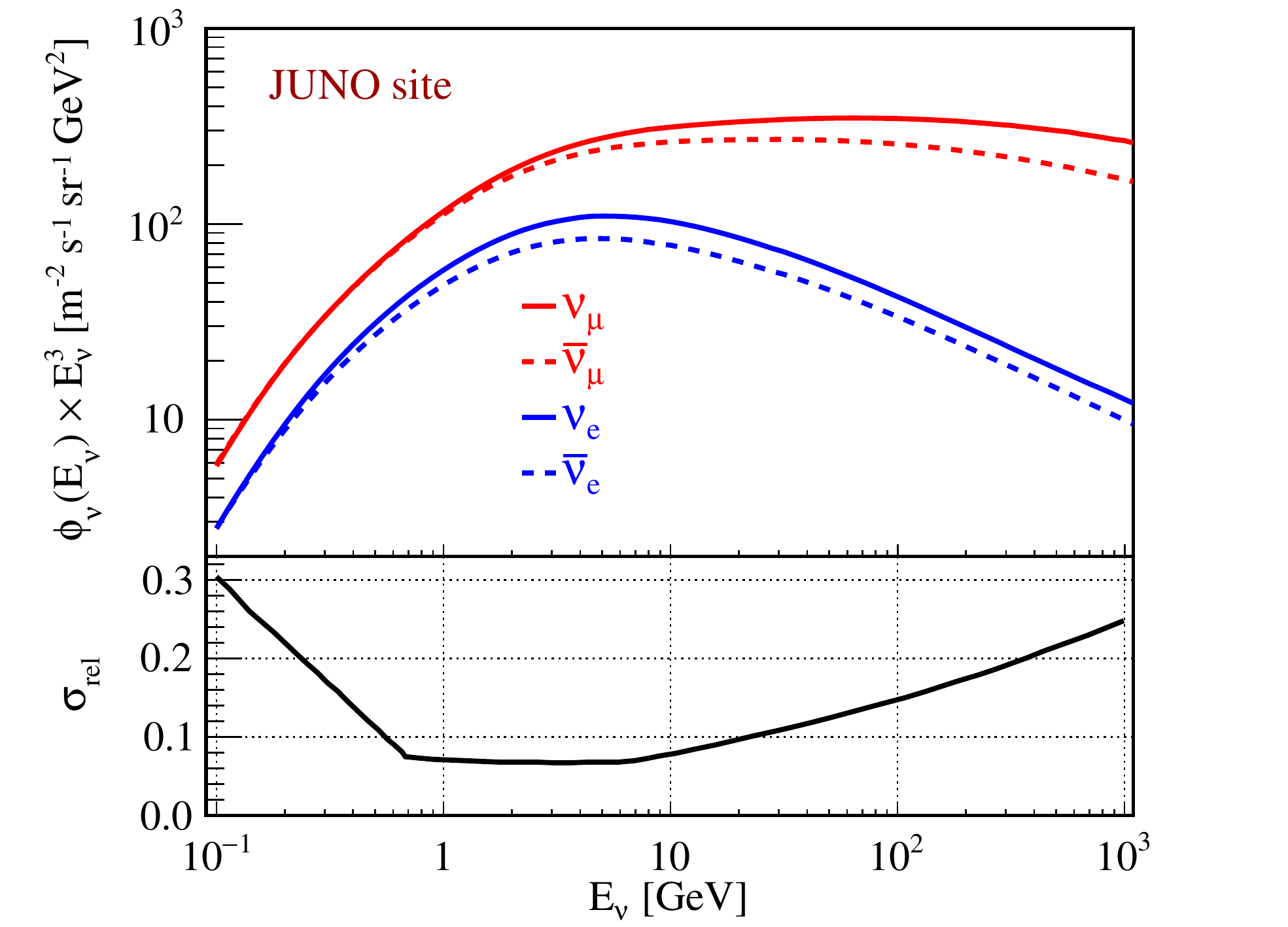}
\end{center}
\vspace{-0.8cm}
\caption{{The predicted fluxes $\phi^{}_\nu(E^{}_\nu) \times E^3_\nu~[{\rm m}^{-2}~{\rm s}^{-1}~{\rm sr}^{-1}~{\rm GeV}^2]$ (upper panel) and the associated relative uncertainty (lower panel) $\sigma_{\rm rel}$ } of atmospheric neutrinos for $\nu = \nu^{}_\mu, \overline{\nu}^{}_\mu, \nu^{}_e, \overline{\nu}^{}_e$ at the JUNO site, as calculated by the Honda group in Ref.~\cite{Hondahomepage}. 
\label{fig:flux}}
\end{figure}

{In the upper panel of} Fig.~\ref{fig:flux}, we show $\phi^{}_\nu(E^{}_\nu) \times E^3_\nu~[{\rm m}^{-2}~{\rm s}^{-1}~{\rm sr}^{-1}~{\rm GeV}^2]$, namely, the predicted fluxes $\phi^{}_\nu(E^{}_\nu)$ of atmospheric neutrinos multiplied by $E^3_\nu$ for $\nu = \nu^{}_\mu, \overline{\nu}^{}_\mu, \nu^{}_e, \overline{\nu}^{}_e$ , at the JUNO site, as calculated by the Honda group~\cite{Hondahomepage}. Notice that the fluxes {shown here} have been averaged over all directions, and there is no mountain assumed at the JUNO site~\footnote{The mountain profile may have large effects on the atmospheric neutrino flux below 100 MeV~\cite{Guo:2018sno}. A careful calculation at the JUNO site with accurate mountain profile is still ongoing.}. In addition, the neutrino energies are ranging from $100~{\rm MeV}$ to $10^{4}~{\rm GeV}$. {The uncertainty in the predictions for atmospheric neutrino fluxes is illustrated in the lower panel of Fig.~\ref{fig:flux}, which varies from about $30\%$ at low energies to less than $10\%$ in the range of $(1-10)~{\rm GeV}$.} Starting from the neutrino energy around $10^{}~{\rm GeV}$, the uncertainty is gradually increasing mainly from the variations of hadron interaction models for $\pi$ and $K$ production.
{As already pointed out in Refs.~\cite{Honda:2006qj,Honda:2019ymh},} the uncertainty of neutrino fluxes could be further reduced by more precise measurements of cosmic muon flux at the same experimental site.

\subsection{Neutrino Interaction Generators}\label{subsec:generator}

\begin{figure}[!t]
\begin{center}
\begin{tabular}{l}
\hspace{-0.6cm}
\includegraphics[width=0.8\textwidth]{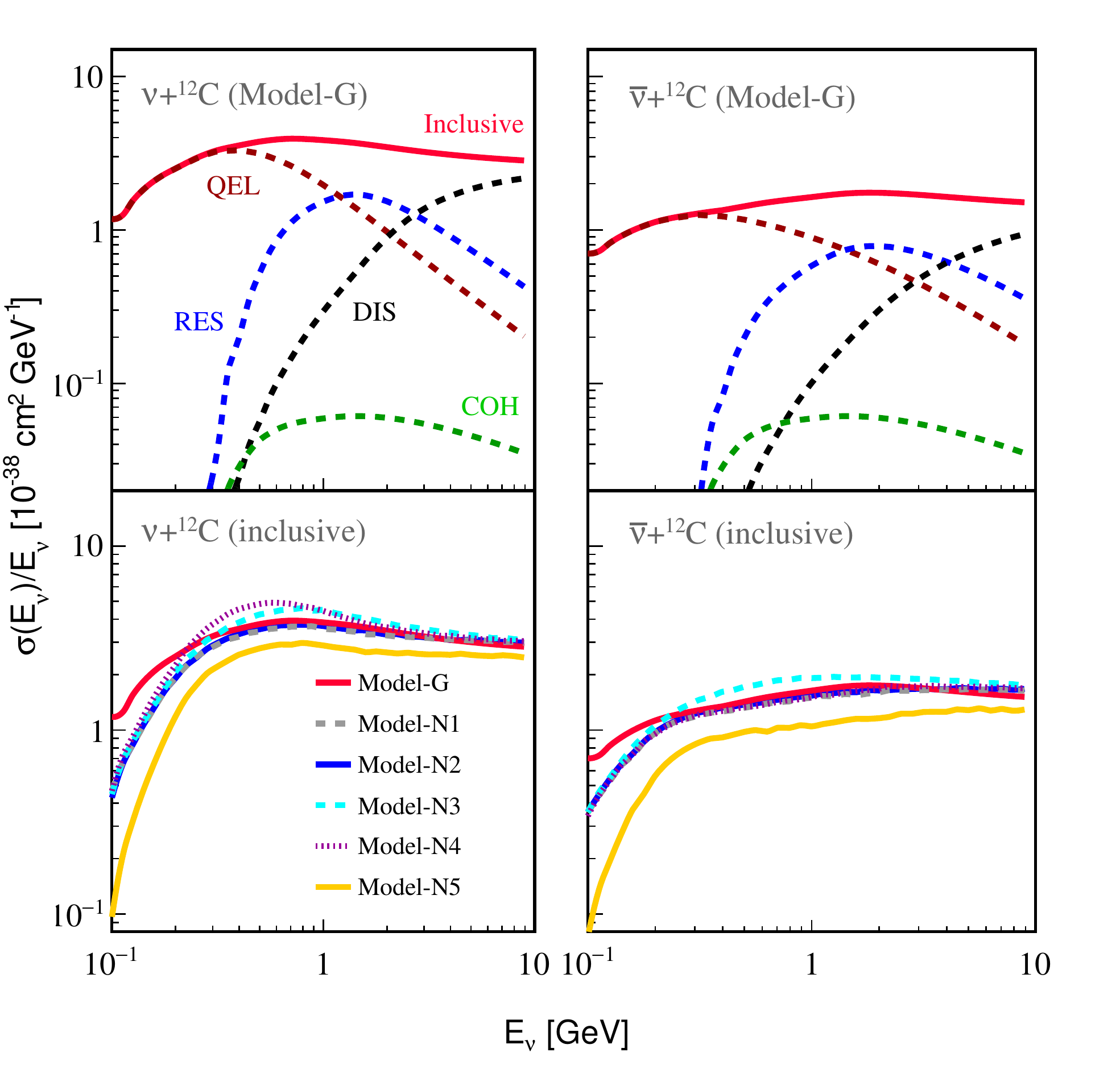}
\end{tabular}
\end{center}
\vspace{-0.8cm}
\caption{The cross section $\sigma(E^{}_\nu)/E^{}_\nu~[10^{-38}~{\rm cm}^2~{\rm GeV}^{-1}]$ of the neutral-current (NC) interactions between neutrinos and the $^{12}$C nucleus has been given in the left panel, while that for antineutrinos $\sigma(E^{}_{\overline{\nu}})/E^{}_{\overline{\nu}}~[10^{-38}~{\rm cm}^2~{\rm GeV}^{-1}]$ in the right panel, where the results for all six typical models in Table~\ref{table:model} have been shown.
\label{fig:cross}}
\end{figure}

Next, given the fluxes of atmospheric neutrinos, we proceed {by simulating} the neutrino interactions with $^{12}{\rm C}$ in the LS detectors. {Our calculation of the} cross section for the NC interactions between neutrinos and nuclei suffers from various uncertainties mainly in the nuclear structure and many-body effects in the nuclei. First, the individual nucleon participating in the NC interactions is actually confined in the nucleus, so the nuclear structure of the latter should be taken into account. Second, as the energy of the incident neutrino increases from $100~{\rm MeV}$ to GeV or even higher, the dominant contribution to the cross section comes roughly from quasi-elastic scattering (QEL), coherent and diffractive production (COH), nuclear resonance production (RES) and deep inelastic scattering (DIS) in different energy ranges. In general, all these processes should be considered.
{Taking the generator GENIE as an example, the transition from QEL to COH and RES to DIS is shown in the upper panel of Fig.~\ref{fig:cross} for both neutrino (left) and antineutrino (right) cross sections. The inclusive cross sections are also provided as the combination of all above contributions.}
Third, the nuclear effects, such as the final-state interaction, the meson exchange current (MEC) and multi-nucleon correlation, must be included as well in a more complete study~\cite{LlewellynSmith:1971uhs, Donnelly:1978tz, Formaggio:2013kya}.

\begin{table}
\setlength{\tabcolsep}{9pt}
\centering
\begin{tabular}{ccccc}
\hline
\multicolumn{1}{c}{Models}  & \multicolumn{1}{c}{Generator} &  \multicolumn{1}{c}{$M^{}_{\rm A}~[{\rm GeV}]$}  & \multicolumn{1}{c}{Nuclear Model}  & \multicolumn{1}{c}{TEM} \\ 
\hline
\hline
\multicolumn{1}{l}{Model-G} & \multicolumn{1}{c}{\texttt{GENIE}} & \multicolumn{1}{c}{0.99}  & \multicolumn{1}{c}{RFG}  & \multicolumn{1}{c}{No}\\ 
\multicolumn{1}{l}{Model-N1} & \multicolumn{1}{c}{\texttt{NuWro}} &\multicolumn{1}{c}{0.99}  & \multicolumn{1}{c}{RFG}  & \multicolumn{1}{c}{No}\\ 
\multicolumn{1}{l}{Model-N2} & \multicolumn{1}{c}{\texttt{NuWro}} &\multicolumn{1}{c}{1.03}  & \multicolumn{1}{c}{RFG}  & \multicolumn{1}{c}{No}\\ 
\multicolumn{1}{l}{Model-N3} & \multicolumn{1}{c}{\texttt{NuWro}} &\multicolumn{1}{c}{1.35}  & \multicolumn{1}{c}{RFG}  & \multicolumn{1}{c}{No}\\ 
\multicolumn{1}{l}{Model-N4} & \multicolumn{1}{c}{\texttt{NuWro}} &\multicolumn{1}{c}{0.99}  & \multicolumn{1}{c}{RFG}  & \multicolumn{1}{c}{Yes}\\ 
\multicolumn{1}{l}{Model-N5} & \multicolumn{1}{c}{\texttt{NuWro}} &\multicolumn{1}{c}{0.99}  & \multicolumn{1}{c}{SF}  & \multicolumn{1}{c}{No}\\ 
\hline
\end{tabular}
\vspace{0.3cm}
\caption{Summary of the main features of six different models extracted from the Monte Carlo generators of neutrino interactions \texttt{GENIE}~\cite{Andreopoulos:2009rq} and \texttt{NuWro}~\cite{Golan:2012rfa}, where the adopted models differ in the input values of the axial mass $M^{}_{\rm A}$, whether the relativistic Fermi gas (RFG) model of or the spectral function (SF) approach to nuclear structure is used, or whether the transverse enhancement model (TEM) in the two-body current contribution is considered.}
\label{table:model}
\end{table}
In order to take account of different contributions and handle nuclear effects in a proper way in neutrino interactions with nuclei, one usually relies on sophisticated Monte Carlo {neutrino event} generators, which have been carefully constructed and finely adjusted according to the available experimental data on the interaction cross sections. In addition to \texttt{GENIE} and \texttt{NuWro}, which will be used in our calculations to illustrate the dependence on nuclear models, other generators \texttt{NUANCE}~\cite{Casper:2002sd}, \texttt{NEUT}~\cite{Hayato:2009zz} and \texttt{GiBUU}~\cite{Buss:2011mx} are available. See, e.g., Ref.~\cite{Mosel:2019vhx}, for a recent review on the neutrino event generators. Depending on incident neutrino energies, the NC interactions with the $^{12}$C nuclei will be dominated by the QEL for the energy range of several hundred MeV, or by the DIS for energies above 100 GeV. For neutrino energies in between, the NC interactions will be complicated by multiple production processes and various nuclear effects~\cite{Formaggio:2013kya}. In the present work, we utilize the software version of \texttt{GENIE (2.12.0)} and \texttt{NuWro {(17.01)}} to calculate the cross sections. Both generators provide some options for the physics parameters and nuclear models, among which one key parameter is the axial mass $M^{}_{\rm A}$ in the parametrization of the nuclear axial-vector form factor. In \texttt{GENIE}, the default setting is $M^{}_{\rm A} = 0.99~{\rm GeV}$ {for the QEL process}, which has been determined from the deuterium measurements~\cite{Kitagaki:1990vs}. In \texttt{NuWro}, this parameter is tunable, and thus we take the following three different values.
\begin{itemize}
\item  $M^{}_{\rm A}= 0.99\,{\rm GeV}$, i.e., identical to that in \texttt{GENIE}, in order to study the systematic differences arising from the modelling in \texttt{GENIE} and \texttt{NuWro}.

\item  $M^{}_{\rm A}= 1.35\,{\rm GeV}$, which is mainly motivated by the latest analysis of the MiniBooNE data on neutrino-carbon interactions~\cite{AguilarArevalo:2010zc}. The MiniBooNE data indicate a surprisingly large cross section, which has triggered an intense discussion on possible explanations~\cite{Morfin:2012kn, Alvarez-Ruso:2017oui}, {where the existence of a MEC component is the leading explanation of this discrepancy.}

\item  $M^{}_{\rm A}= 1.03\,{\rm GeV}$, which is actually taken the world average from Ref.~\cite{Bernard:2001rs}. Since this value is very close to $M^{}_{\rm A} = 0.99~{\rm GeV}$, the final results are expected to be rather similar.
\end{itemize}

Regarding the models of nuclear structure, \texttt{GENIE} uses the relativistic Fermi gas (RFG) model {with `Bodek-Ritchie' modifications~\cite{Bodek:1981wr}}~\footnote{{The `Bodek-Ritchie' modifications~\cite{Bodek:1981wr} add a high momentum component to emulate the high momentum observed in nuclear response data (and also included in the Benhar Spectral Function).}} as a default setting, while
\texttt{NuWro} provides several choices for the description of the target nucleus, including both RFG and the spectral function (SF) approach.
Furthermore, to illustrate the two-body current effects in QEL, {we also take the transverse enhancement model (TEM) of MEC from \texttt{NuWro}}, which has been obtained in Ref.~\cite{Bodek:2011ps} by fitting the electron scattering data. By including TEM, the authors of Ref.~\cite{Bodek:2011ps} have demonstrated that the MiniBooNE results can be reproduced with a smaller value of the axial mass, which turns out to be consistent with the world average value obtained from other measurements. It should be emphasized that only the axial mass $M^{}_{\rm A}$ in the treatment of QEL in \texttt{NuWro} has been changed. For both generators, we shall employ their default setting for all other processes.

In Table~\ref{table:model}, we summarize the main features of six typical models, which have been implemented in our calculation to illustrate the model variations of neutrino interactions. Some comments on these models are in order. Just one model from \texttt{GENIE} (i.e., Model-G) has been adopted, while Model-N1 from \texttt{NuWro} has been selected with the same input for the nuclear model and the axial mass. The motivation for this setup is to make a comparison between \texttt{GENIE} and \texttt{NuWro}. Within the same generator \texttt{NuWro}, Model-N1, Model-N2 and Model-N3 share exactly the same setup, except for different values of the axial mass $M^{}_{\rm A}$ for the QEL process, which is intended to illustrate the impact of the axial mass on the NC cross section. Additionally, two other models in \texttt{NuWro} are introduced. First, Model-N4 is the only one to include TEM, so that we can examine the two-body current effect by comparing between Model-N1 and Model-N4. Second, the SF approach is incorporated in Model-N5, whereas the RFG model is used in all others, offering {the} possibility to study the difference between these two nuclear models within \texttt{NuWro}.
{Note that current model variations employed here are still limited. A comprehensive set of up-to-date neutrino event generators, including all aforementioned ones, will be required to make the reliable uncertainty evaluations.
}

{In the lower panel of Fig.~\ref{fig:cross}}, we have illustrated the inclusive cross sections for the NC interactions of neutrinos and antineutrinos with $^{12}{\rm C}$ from \texttt{GENIE} and \texttt{NuWro}, and shown the final results of $\sigma(E^{}_\nu)/E^{}_\nu$ and $\sigma(E^{}_{\overline{\nu}})/E^{}_{\overline{\nu}}$ for six representative models in Table~\ref{table:model}. A number of important observations can be made {in this respect}.

First, for both neutrino and antineutrino cross sections, Model-N5 gives the smallest value of the inclusive cross section. This is the only model that uses the SF approach for the target nucleus, instead of the RFG model. As previously shown in Refs.~\cite{Benhar:2006nr, Ankowski:2007uy}, the inclusion of many-body nuclear effects in the SF approach generally reduces the total cross sections by $20\%$ or so at the energy around $1~{\rm GeV}$. This effect can be clearly seen by comparing between the cross section at $E^{}_\nu = 1~{\rm GeV}$ (or $E^{}_{\overline{\nu}} = 1~{\rm GeV}$) in Model-N1 and that in Model-N5, for which the same axial mass $M^{}_{\rm A} = 0.99~{\rm GeV}$ is input. For even lower energies, the relative difference between these two models becomes more significant. However, the impulse approximation made in {the RFG and SF approaches} will be invalidated for low momentum transfers~\cite{Benhar:2006nr}, rendering such a comparison to be problematic. Second, the predictions from Model-N1 and Model-G are well consistent with each other for the energies above $300~{\rm MeV}$, but remarkable deviations can be found for lower energies. This discrepancy might be attributed to the different treatments of nuclear effects in \texttt{NuWro} and \texttt{GENIE},
and will be taken as a systematic uncertainty in our calculations. Third, the axial mass $M^{}_{\rm A} = 1.35~{\rm GeV}$ is set in Model-N3, leading to the largest cross section for both neutrinos and antineutrinos. {However, Model-N4 with $M^{}_{\rm A} = 0.99~{\rm GeV}$ predicts even larger cross section below $1~{\rm GeV}$ due to TEM.} Both of them are motivated by the MiniBooNE data, which does not necessarily mean more accurate description of microscopic physics. Notice that the TEM effect is absent in the antineutrino sector, so the predicted cross sections in Model-N1, Model-N2 and Model-N4 essentially coincide. Finally, it is evident that the cross section for neutrinos turns out to be larger than that for antineutrinos in all models under consideration. The main reason is that the opposite sign of the axial-vector coupling for neutrinos and antineutrinos, which have the opposite helicities, results in the constructive interference in the transverse- and axial-vector amplitudes for neutrinos but the destructive interference for antineutrinos~\cite{Gallagher:2011zza, Mosel:2016cwa}.

\subsection{Deexcitation of Final-state Nuclei}
\label{subsec:deexc}

\begin{table}[!t]
\setlength{\tabcolsep}{9pt}
\centering
\begin{tabular}{cccc}
\hline
\multicolumn{1}{c}{\multirow{1}{*}{Daughter Nuclei}} & \multicolumn{1}{c}{\multirow{1}{*}{ Shell Hole}}  & \multicolumn{1}{c}{Configuration Probability}  & \multicolumn{1}{c}{{Excitation Energy}} \\
\hline
\multicolumn{1}{c}{\multirow{2}{*}{$^{11}$C$^*$\;\rm {or}\;$^{11}$B$^*$}} & \multicolumn{1}{c}{$s_{1/2}$} & $1/3$  & \multicolumn{1}{c} {$E^* = 23\,{\rm MeV}$} \\
& \multicolumn{1}{c}{$p_{3/2}$} & $2/3$& \multicolumn{1}{c} {$E^* = 0\,{\rm MeV}$} \\
\hline
\multicolumn{1}{c}{\multirow{3}{*}{$^{10}$C$^{*}$\;\rm {or}\;$^{10}$Be$^{*}$}} & \multicolumn{1}{c}{$s_{1/2}$} & $1/15$  & \multicolumn{1}{c} {$E^* = 46\,{\rm MeV}$} \\
& \multicolumn{1}{c}{$p_{3/2}$} & $6/15$ & \multicolumn{1}{c} {$E^{*} = 0\,{\rm MeV}$} \\
& \multicolumn{1}{c}{$s_{1/2}$\,\&\,$p_{3/2}$} & $8/15$   & \multicolumn{1}{c} {$E^* = 23\,{\rm MeV}$}  \\

\hline
\multicolumn{1}{c}{\multirow{3}{*}{$^{10}$B$^{*}$}}
& \multicolumn{1}{c}{$s_{1/2}$} & $1/7$  & \multicolumn{1}{c} {$E^{*} = 46\,{\rm MeV}$} \\
& \multicolumn{1}{c}{$p_{3/2}$} & $4/7$  & \multicolumn{1}{c} {$E^{*} = 0\,{\rm MeV}$} \\

& \multicolumn{1}{c}{$s_{1/2}$\,\&\,$p_{3/2}$} & $2/7$   & \multicolumn{1}{c} {$E^* = 23\,{\rm MeV}$} \\
\hline
\end{tabular}
\vspace{0.3cm}
\caption{The probabilities of the configurations for the nuclei $^{11}{\rm C}$ and $^{11}{\rm B}$ with one nucleon disappearing from ${^{12}\textrm{C}}$ in the nuclear model and those for $^{10}{\rm C}$, $^{10}{\rm Be}$ and $^{10}{\rm B}$ with two nucleons less, where the corresponding excitation energies $E^*$ are given in the last column and $E^* = 0~{\rm MeV}$ actually refers to the ground state. }
\label{table:prob}
\end{table}
The last step is to deal with the deexcitation of final-state nuclei from the NC interactions and extract the total energy spectra of $\gamma$'s, light mesons, protons, neutrons, and $\alpha$'s, which may contribute to the irreducible backgrounds for the experimental searches for rare events in the LS detectors. However, the Monte Carlo {event} generators of neutrino interactions usually do not provide the exact state of the residual nucleus, which may reside in one of its various excited states,
and the nuclear deexcitation with additional $\gamma$ rays, protons, neutrons or other heavier projectiles is important. Thus it is desirable to combine the neutrino interaction generators with the nuclear structure model and the deexcitation tool to complete our calculations.

To this end, the widely-used \texttt{TALYS} software~\cite{Koning:2005ezu} will be employed to treat the deexcitation of daughter nuclei. In general, \texttt{TALYS} is a very useful tool for both nuclear structures and nuclear reactions in the energy range from $1\,{\rm keV}$ to $200\,{\rm MeV}$. For our purpose, the branching fractions of the deexcitation of the excited nucleus and the energy spectra of all the relevant products from the deexcitation are needed. More explicitly, after the excited state of a nucleus $N^*$ with the excitation energy $E^*$ is chosen, we implement \texttt{TALYS} to simulate all possible deexcitation channels of the given nucleus and follow the further deexcitation of the residual nuclei until all the final-state particles are in their ground states, either stable or beta decaying. When the residual nuclei are unstable, we {simulate} their decay processes based on the decay types, endpoints, and lifetimes from the nuclear database. Hence the energy spectra of all deexcitation products are obtainable.

Before doing so, we have to specify the nuclear structure of the target nucleus $^{12}{\rm C}$. In order to simplify our discussions, we make use of the statistical configuration from the nuclear shell model of $^{12}$C~\cite{Kamyshkov:2002wp,Auerbach:1997ay,Kolbe:1999au}. Since there are six protons and six neutrons in the $^{12}{\rm C}$ nucleus, the $s^{}_{1/2}$ and $p^{}_{3/2}$ shells of the lowest energy level for both protons and neutrons will be occupied in the simplest configuration. Moreover, we neglect the potential configuration of the $^{12}\rm{C}$ nucleus due to the nucleon pairing correlation between the $p^{}_{3/2}$ and $p^{}_{1/2}$ shells because of the relatively small energy gap of around 3-4 MeV~\cite{Kamyshkov:2002wp,Auerbach:1997ay}. This effect will be included in our future work with more sophisticated shell model calculations.

Using this statistical model of the ground state of $^{12}$C, we can figure out all possible excited states for those lighter daughter nuclei resulting from the neutrino interactions with $^{12}$C. The excited states of daughter nuclei are obtained by considering the disappearance of one or more nucleons (either protons or neutrons) from the $^{12}$C ground state. For instance, we show the cases of one or two nucleons disappearing from $^{12}{\rm C}$ in Table~\ref{table:prob}, where the corresponding probabilities of possible configurations are provided together with the excitation energies of these daughter nuclei. The information of these daughter nuclei will be further input in \texttt{TALYS} to obtain their deexcitation and the energy spectra of the associated final-state particles. In a similar way, one can discuss other possibilities of more nucleons disappearing from $^{12}{\rm C}$. In our calculations, all the daughter nuclei with a mass number larger than five have been taken into account.

\section{Results and Discussions}\label{sec:results}

Sec.~\ref{sec:framework} describes the ingredients necessary for a numerical calculation of the background induced by NC interactions of atmospheric neutrinos: the atmospheric neutrino fluxes at the JUNO site in Fig.~\ref{fig:flux}, the inclusive cross sections of (anti)neutrino-$^{12}$C interactions in Fig.~\ref{fig:cross}, the possible excited states of daughter nuclei (e.g., those in Table~\ref{table:prob}) and the corresponding deexcitation processes.
In this Section, we proceed to carry out a detailed analysis of the NC backgrounds for the detection of DSNB and proton decay.
We shall first present the general calculation of NC interactions in Sec.~\ref{subsec:general}, and then the specific results on the background in searches of
DSNB and proton decay in Sec.~\ref{subsec:dsnb} and Sec.~\ref{subsec:pd} respectively.
However, it is worthwhile {emphasizing} that the background analysis for any practical search of rare events depends very much on the properties of signals, the performance of the LS detectors and other advanced techniques. Thus we concentrate on the main features of the NC backgrounds and leave the intricate strategy for background reduction for future and better works by the experimental collaborations.

\subsection{NC interaction rates}\label{subsec:general}

We calculate the event rate of atmospheric neutrino NC interactions as a function of the energy of the incident neutrino or antineutrino based on
\begin{equation}
\label{ER}
n(E^{}_{\nu})= 4 \pi N \sum_{\nu} \phi^{}_{\nu}(E^{}_{\nu}) \times \sigma^{}_{\nu}(E^{}_{\nu}) ,
\end{equation}
where $\nu = \nu^{}_\mu, \overline{\nu}^{}_\mu, \nu^{}_e, \overline{\nu}^{}_e$ runs over neutrinos and antineutrinos of both electron and muon flavors, the factor $4\pi$ comes from the integration over the full solid angle, and $ N \approx 4.4 \times 10^{31}$ is the total number of target ${^{12}\textrm{C}}$ nuclei per kiloton LS, where the contribution of ${^{13}\textrm{C}}$ at the level of 1\% has been neglected for simplicity. {The cross section of atmospheric neutrino NC interactions on Hydrogen is one order of magnitude smaller than that of ${^{12}\textrm{C}}$, and that with one or more neutron production is negligible.}
Note that the cross section $\sigma^{}_{\nu}(E^{}_\nu)$ could be either the inclusive one as shown in Fig.~\ref{fig:cross} or the exclusive one with specified final states {discussed later.} 
\begin{figure}[!t]
\begin{center}
\includegraphics[width=0.7\textwidth]{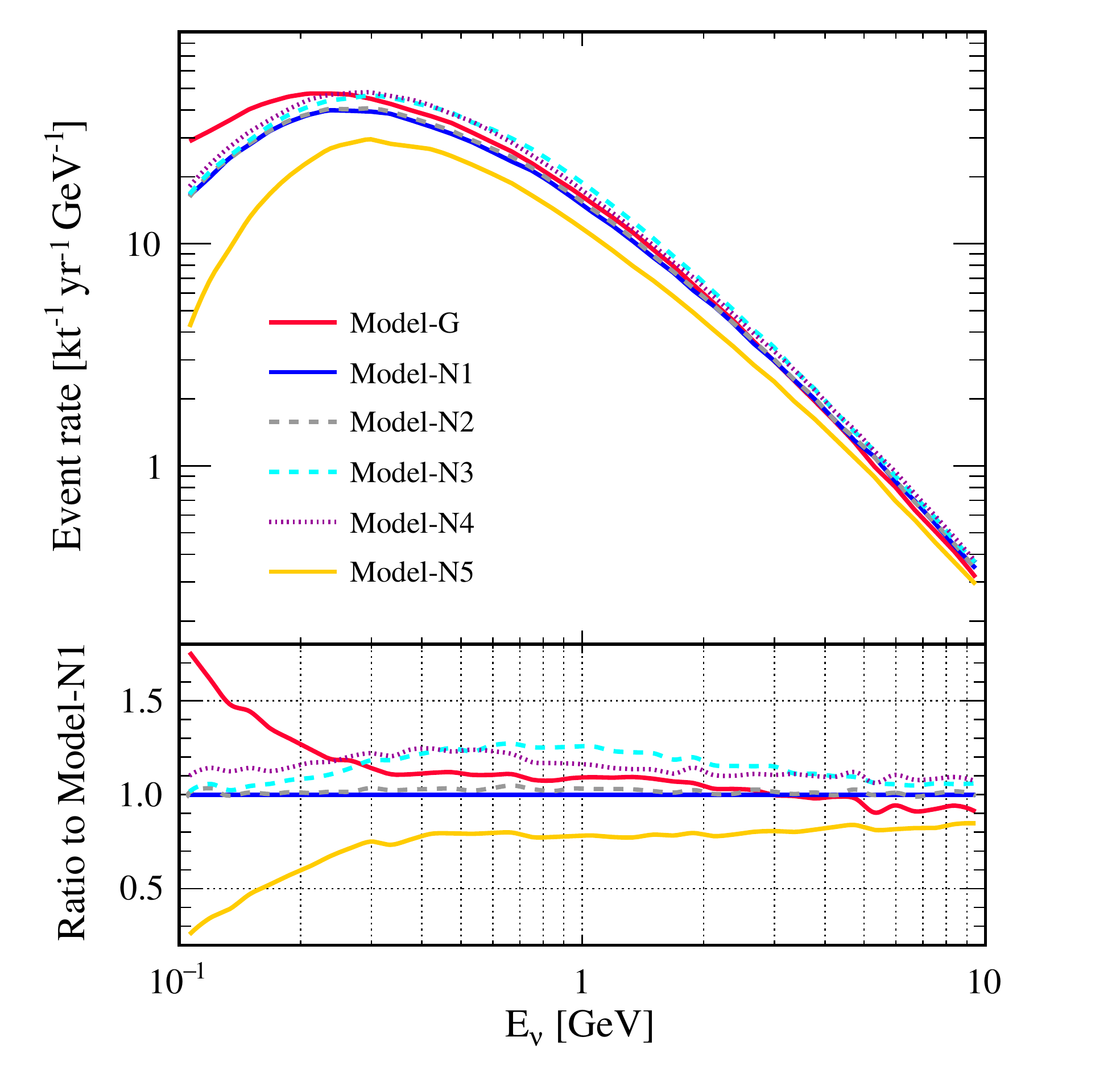}
\end{center}
\vspace{-0.8cm}
\caption{The distribution of the total event rate induced by atmospheric neutrino NC interactions with respect to the neutrino energy for each of six representative interaction models in Table~\ref{table:model}. The atmospheric neutrino fluxes in Fig.~\ref{fig:flux} and the inclusive cross sections in Fig.~\ref{fig:cross} have been input in the calculations, and the ratio of the event rate in each model to that in Model-N1 has been shown in the lower panel.
\label{fig:eventrate}}
\end{figure}

The distribution of the total event rate with respect to the neutrino energy has been calculated by convolving the atmospheric (anti)neutrino fluxes in Fig.~\ref{fig:flux} and the inclusive cross sections in Fig.~\ref{fig:cross} and by summing up {all the neutrino species}. The final results for the six representative models in Table~\ref{table:model} are shown in Fig.~\ref{fig:eventrate}. For comparison, in the lower panel of Fig.~\ref{fig:eventrate}, the ratio of the result in each model to that in Model-N1 is presented. 

In addition, we have checked the individual contributions from different physical processes (i.e., QEL, RES and DIS) to the inclusive event rates in six representative models. Roughly speaking, QEL, RES and DIS respectively contribute around 60\%, 20\%, 10\% of the total event rate. QEL is the dominant process in the energy range of the DSNB signal, while RES and DIS are the major processes in the range for the proton decay search.
{Note that the inclusion of TEM in Model-N4} leads to a remarkable impact, which compensates the relatively smaller event rate of QEL in this model due to a smaller value of $M^{}_{\rm A}$. On the other hand, the individual contributions from $\nu^{}_{\mu}$, $\nu^{}_{e}$, $\overline{\nu}^{}_{\mu}$, $\overline{\nu}^{}_{e}$ {account respectively for about $50\%$, $20\%$, $20\%$ and $10\%$ of the total events in all six models.}
This finding is directly related to the differences in the fluxes and the inclusive cross sections. Firstly, the inclusive cross section for antineutrinos is smaller than that for neutrinos, as shown in Fig.~\ref{fig:cross}. Secondly, the atmospheric neutrino flux of the muon flavor is dominant.
\begin{figure}[!t]
\begin{center}
\begin{tabular}{l}
\includegraphics[width=1.0\textwidth]{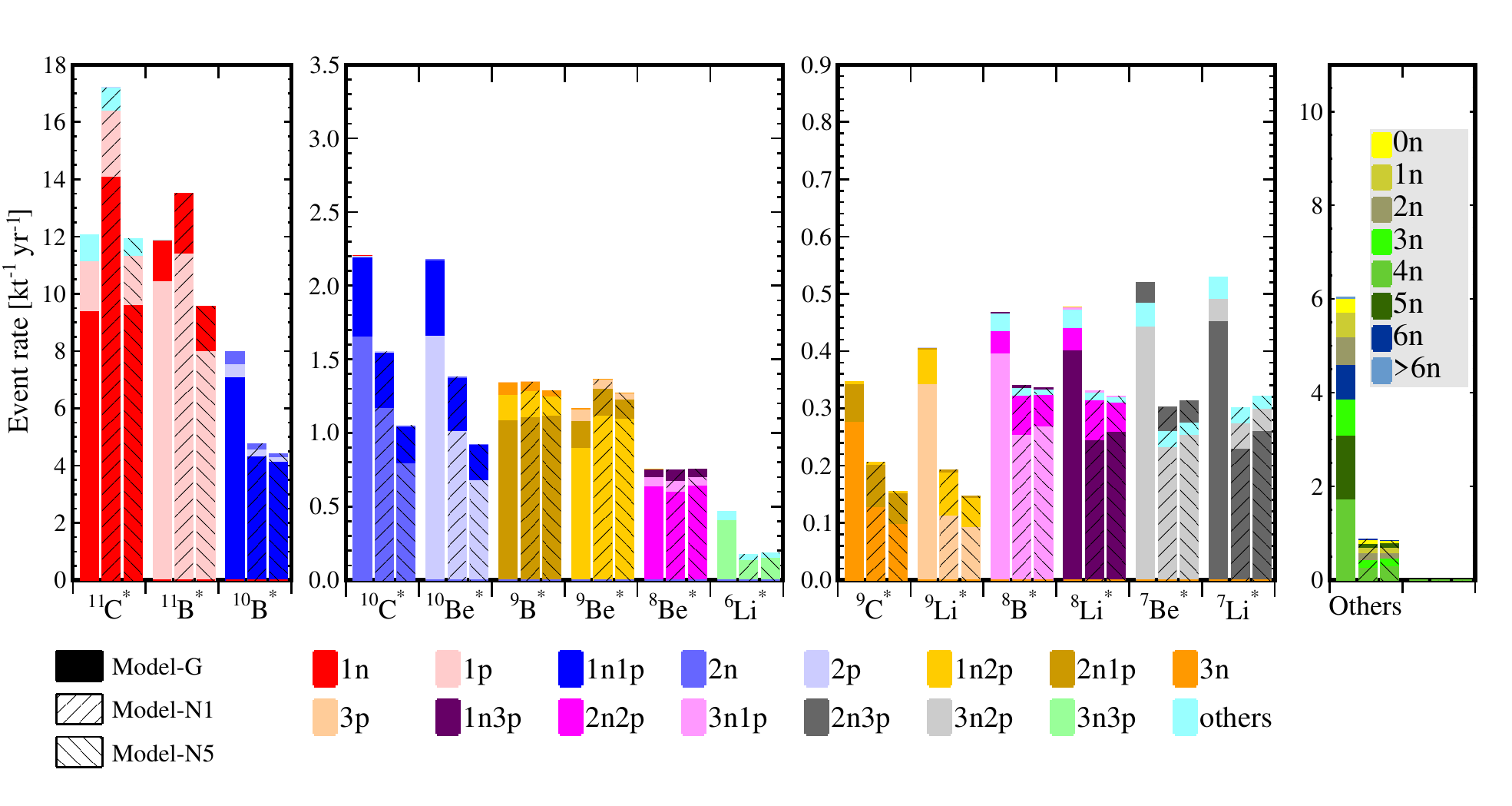}
\end{tabular}
\end{center}
\vspace{-1.0cm}
\caption{The event rates for the NC interactions of atmospheric neutrinos (with energies ranging from $100~{\rm MeV}$ to $10~{\rm GeV}$) with $^{12}$C nuclei in the exclusive channels, which have been categorized by the daughter residual nuclei and the associated superscript ``$\ast$" {highlights} possible excited states. In each channel, the event rates predicted by Model-G, Model-N1 and Model-N5 are shown as the solid filled histogram, the striped histogram with slashes, and the striped histogram with backslashes, respectively. For each exclusive channel in a specified model, the contributions from the processes with one or more extra nucleons are represented by the colored bars. The results for $^{12}{\rm C}$ and other isotopes with the atomic number $Z < 3$ or the mass number $A < 6$ have been given in the rightmost panel and labelled by ``Others".
\label{fig:table1}}
\end{figure}
\begin{figure}[!t]
\begin{center}
\begin{tabular}{l}
\includegraphics[width=1.0\textwidth]{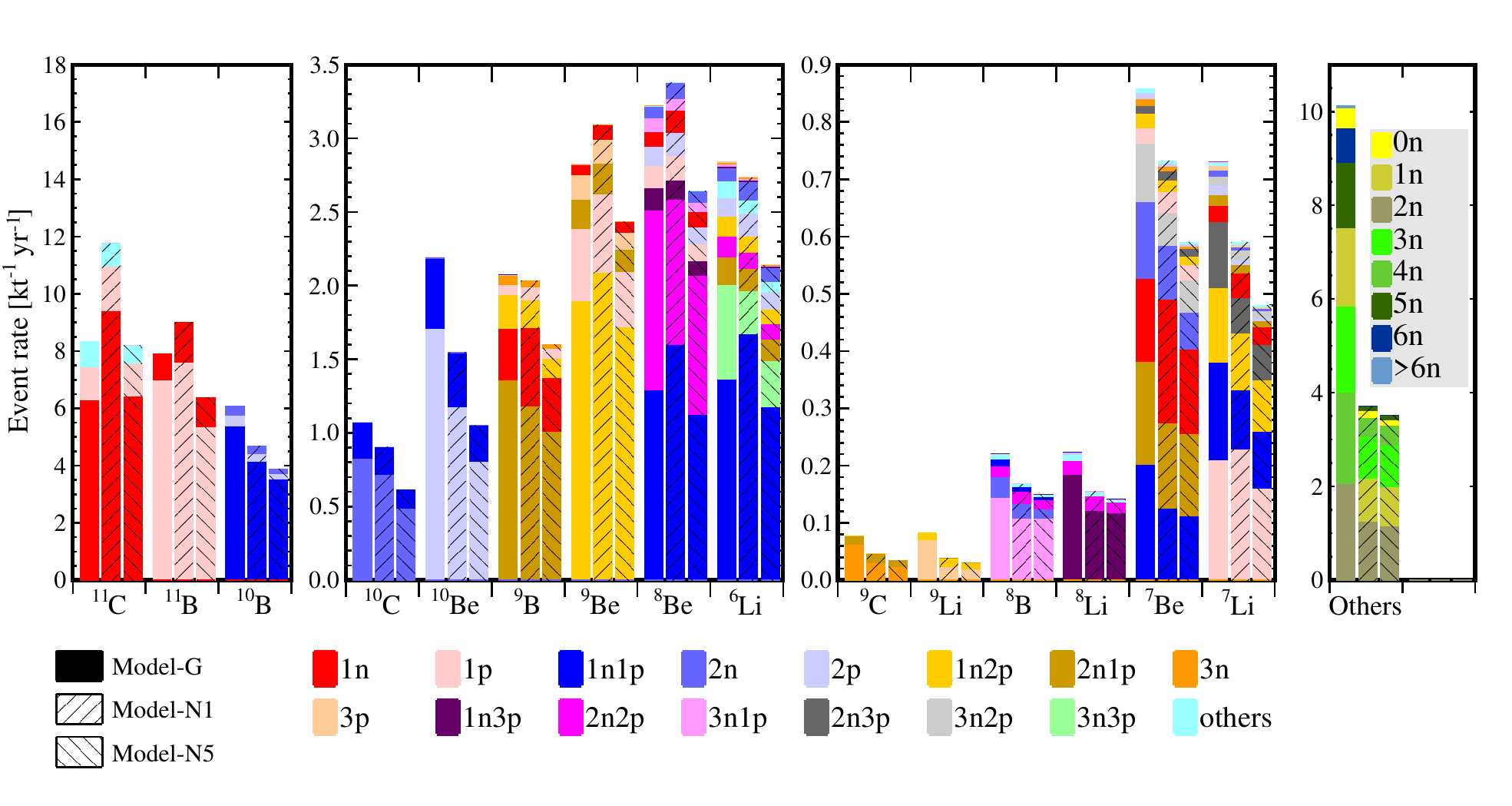}
\end{tabular}
\end{center}
\vspace{-1.0cm}
\caption{The event rates for the NC interactions of atmospheric neutrinos (with energies ranging from $100~{\rm MeV}$ to $10~{\rm GeV}$) with $^{12}$C nuclei in the exclusive channels, which have been categorized by the daughter residual nuclei in their ground states and the deexcitation of the excited residual nuclei has been simulated by using \texttt{TALYS}. The notations and the patterns of the histograms in each panel exactly follow those in Fig.~\ref{fig:table1}.
\label{fig:table2}}
\end{figure}

In order to examine how the NC interactions of atmospheric neutrinos affect the detection of DSNB and proton decay, we have to go one step further and analyze the final-state products. For this purpose, we calculate the event rates of atmospheric neutrino NC interactions on $^{12}$C nuclei in the exclusive channels, where different final-state nuclei and the production of one or more nucleons should be taken into account. Our final results are summarized in Fig.~\ref{fig:table1} and Fig.~\ref{fig:table2}. 

The relevant energy range of atmospheric neutrinos in question is $100~{\rm MeV} < E^{}_\nu < 10~{\rm GeV}$. Although the atmospheric neutrino fluxes $\phi^{}_\nu(E^{}_\nu)$ multiplied by $E^3_\nu$ in Fig.~\ref{fig:flux} are shown up to $E^{}_\nu = 10^4~{\rm GeV}$, the fluxes $\phi^{}_\nu(E^{}_\nu)$ themselves will be highly suppressed at high energies. For the differential event rates of the inclusive process of NC interactions shown in Fig.~\ref{fig:eventrate}, the maximum rates appear at $E^{}_\nu = (200-300)~{\rm MeV}$. Within this energy range, the QEL process is most important and one or more nucleons will be knocked out from the carbon nucleus. The event rates for the same exclusive processes but with one or more pions {that are produced from the final state interactions} turn out to be negligibly small and can be safely ignored.

In Fig.~\ref{fig:table1}, we summarize the event rates of the relevant exclusive processes, which have been categorized by the residual final-state nuclei and the associated superscript ``$\ast$" {highlights} possible excited states of these nuclei. The results after taking account of the subsequent deexcitation are presented in Fig.~\ref{fig:table2}. We divide the daughter residual nuclei into two groups. The first group includes all the isotopes with the atomic number $Z \ge 3$ and the mass number $A \ge 6$ except for $^{12}$C. The event rates for this group have been shown in the first three panels (from left to right) of Fig.~\ref{fig:table1}, where for each isotope the predictions from Model-G, Model-N1 and Model-N5 are represented by solid filled histograms, the striped histograms with slashes, and the striped histograms with backslashes, respectively. As Fig.~\ref{fig:eventrate} indicated in the previous section, the event rates in the series of models using \texttt{NuWro} (i.e., Model-N$i$ for $i = 1, 2, 3, 4$) are quite similar, therefore, only Model-N1 has been chosen as representative for clarity. For a given model, the individual contributions from the subset of processes with accompanying nucleons can be recognized via the colored bars, where the colors refer to different combinations of the neutron and proton multiplicities.
{Since only the electric charge and baryon number are conserved in the interactions, there are many possibilities to have the transition between nucleons, such as the final state interactions of changing the proton to neutron or vice versa, together with a production of charged mesons.}
Notice that the scales of the vertical axes in all the panels are different from each other. The second group contains all the isotopes with $Z < 3$ or $A < 6$ or $^{12}$C, for which the results are given in the rightmost panel labelled by ``Others". For this group, instead of the final-state nuclei, the subsets of processes have been characterized by the neutron multiplicity, which has been represented by different colored bars. As one can observe from Fig.~\ref{fig:table1}, Model-G generally predicts a higher rate for the exclusive processes with extra nucleons. {The reason can be attributed to different configurations of the final state interactions in event generators, where \texttt{GENIE} with more low-energy events tend to have more hadronic rescattering for lower momentum hadrons.} In addition, the event rate is dominated by the processes with the production of $^{11}{\rm C}^*$, $^{11}{\rm B}^*$, $^{10}{\rm B}^*$, $^{10}{\rm C}^*$, $^{10}{\rm Be}^*$, $^{9}{\rm B}^*$ and $^{9}{\rm Be}^*$, whose deexciation will be further processed by using \texttt{TALYS}.

In Fig.~\ref{fig:table2}, we summarize the final event rates in Model-G, Model-N1 and Model-N5, after taking into account the deexcitation of final-state nuclei from \texttt{TALYS}. As we have explained in the previous section, the probabilities and excitation energies of different configurations for a specified nucleus have been extracted from a simple statistical model of $^{12}{\rm C}$, as partially summarized in Table~\ref{table:prob}. With this input information, one can simulate all possible channels of deexcitation via \texttt{TALYS} until the daughter nuclei are in their ground states. Therefore, all the exclusive processes in Fig.~\ref{fig:table2} are now categorized by the final isotopes that are left in the ground states, where the notations and the patterns of the histograms follow exactly those in Fig.~\ref{fig:table1}. Comparing between Fig.~\ref{fig:table1} and Fig.~\ref{fig:table2}, one can see that the event rates for the processes associated with $^{9}{\rm B}$, $^{9}{\rm Be}$, $^{8}{\rm Be}$, $^{7}{\rm Be}$, $^{7}{\rm Li}$ and $^{6}{\rm Li}$ increase significantly, so does that for ``Others". Meanwhile, the event rates for the processes associated with $^{11}$C, $^{11}$B, $^{10}$C are largely reduced because of the further knockout of one or more nucleon in the deexcitation.  Moreover, the combinations of neutron and proton multiplicities related to each of these nuclear isotopes become more complicated due to the deexcitation processes.


\subsection{DSNB}\label{subsec:dsnb}

First, let us consider the signals of DSNB and possible backgrounds induced by the atmospheric neutrino NC interactions in an LS detector. As has been mentioned in Sec.~\ref{sec:intro}, the IBD process $\overline{\nu}^{}_e + p \to e^+ + n$ is the golden channel for the detection of DSNB $\overline{\nu}^{}_e$, for which the time coincidence between the prompt signal of the positron annihilation and the delayed signal of neutron capture helps {greatly reduce} the background. However, the irreducible backgrounds come from reactor $\overline{\nu}^{}_e$ in the low-energy range and atmospheric $\overline{\nu}^{}_e$ in the high-energy range, leaving only a narrow window of visible energies $11~{\rm MeV} \lesssim E^{}_{\rm vis} \lesssim 30~{\rm MeV}$ for the DSNB observation~\cite{An:2015jdp}.

\begin{figure}[!t]
\begin{center}
\includegraphics[width=0.6\textwidth]{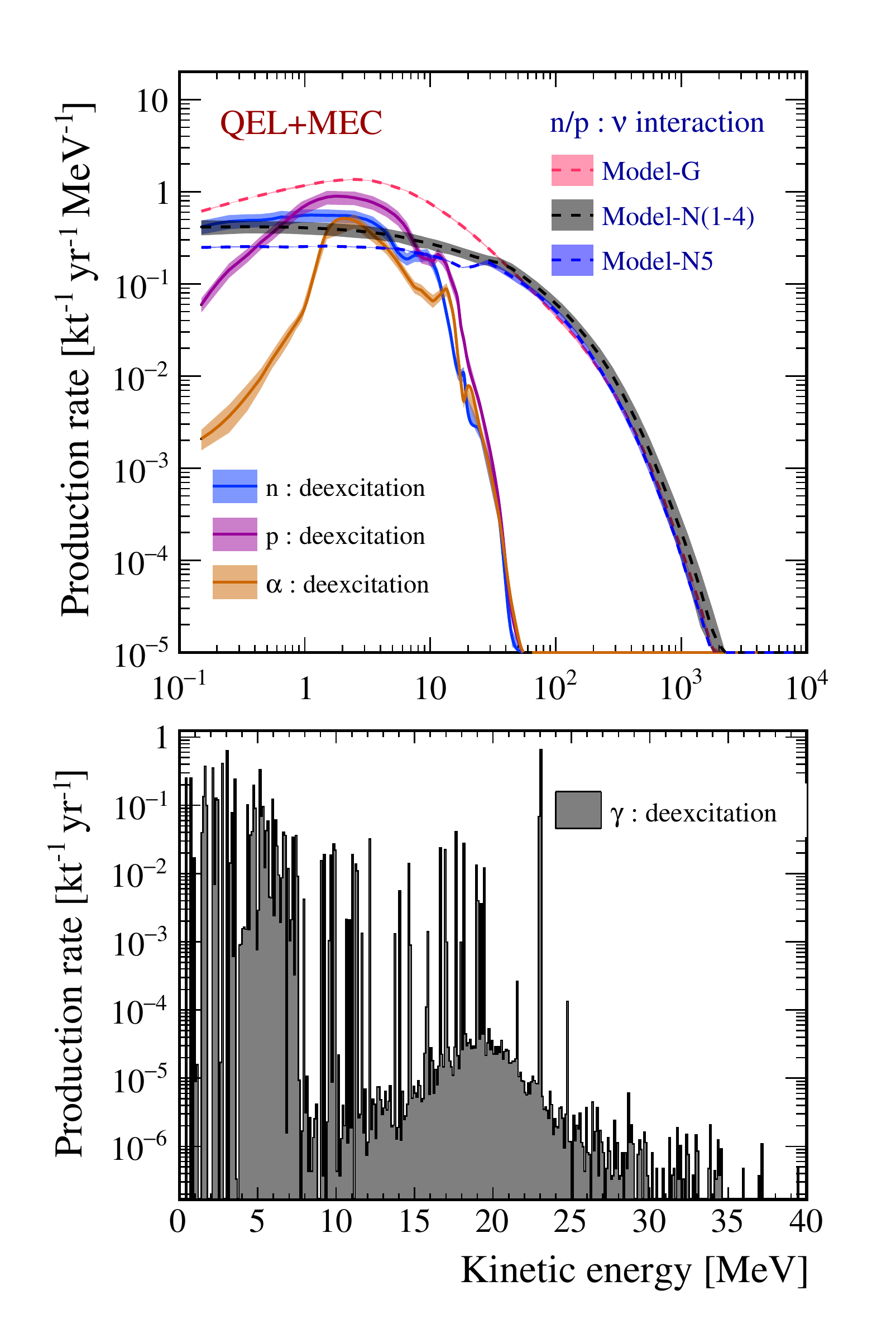}
\end{center}
\vspace{-0.6cm}
\caption{The {kinetic} energy spectra of $p$ or $n$ from the NC interactions of atmospheric neutrinos with $^{12}{\rm C}$ from Model-G (red dashed curve), Model-N$i$ for $i = 1, 2, 3, 4$ (black dashed curve) and Model-N5 (blue dashed curve) have been presented in the upper panel, where those of $p$, $n$ and $\alpha$ from the deexciation of final-state nuclei in the NC interactions have been denoted by blue, magenta and orange solid curves, respectively. The gray band along the black dashed curve represents 1$\sigma$ deviation from the average of the predictions from Model-N$i$ for $i = 1, 2, 3, 4$. The colored band along the solid curve stands for 1$\sigma$ deviation from the average of all six models, for which the deexcitation processes are the same. In the lower panel, the energy spectrum of $\gamma$ from the deexciation of final-state nuclei has been shown as gray histograms. The production rates are shown with respect to the kinetic energies of relevant particles, and the QEL with { MEC} effects dominates the production.
\label{fig:dsnbrate1}}
\end{figure}
Even in this optimal energy window, the NC interactions of atmospheric neutrinos can mimic the DSNB signals. As the energies of our interest are relatively low, the QEL process is of crucial importance. In Fig.~\ref{fig:dsnbrate1}, the {kinetic} energy spectra of $p$, $n$, $\gamma$ and $\alpha$ particles have been extracted from the simulations and are shown with respect to their kinetic energies. In the upper panel, the spectra have been divided into two different categories. First, we show the {kinetic} energy spectra of nucleons, either protons or neutrons, which are directly produced from the NC interactions. In accordance with previous observations, the production rate from the Model-G (red dashed curve) is much higher than those from Model-N$i$ for $i = 1, 2, 3, 4$ (black dashed curve) and Model-N5 (blue dashed curve) below 100 MeV. Since the predictions from Model-N$i$ for $i = 1, 2, 3, 4$ are quite similar, only the average value of them is shown and the gray band along the black dashed curve represents {the standard deviation of different model predictions}. Second, the energy spectra of $p$, $n$ and $\alpha$ particles from the deexcitation of final-state nuclei that are produced in the NC interactions have been plotted as blue, magenta and orange solid curves, respectively. The colored bands along the solid curves stand for {the standard deviation} from the average of all six models. Although the deexcitation processes are the same for these models, the production rates for a given nuclear isotope actually differ. As one can observe from the upper panel, there is significant model dependence of the neutrino interaction generators in the low energy range,
which indicates the necessity of using a complete set of neutrino interaction models.
Meanwhile, the nucleon knockout from the deexcitation of final-state nuclei will be comparable to that directly from NC interactions around or below $10~{\rm MeV}$, but decreases rapidly toward high energies. On the other hand, the energy spectrum of $\gamma$'s has been shown as gray histograms in the lower panel, where discrete lines are superimposed on a spectral continuum reaching to $30-35~{\rm MeV}$. The highest-rate line at $E^{}_\gamma = 23~{\rm MeV}$ can be attributed to the excitation energy of several excited nuclei in Table~\ref{table:prob}.

\begin{figure}[!thb]
\begin{center}
\includegraphics[width=0.7\textwidth]{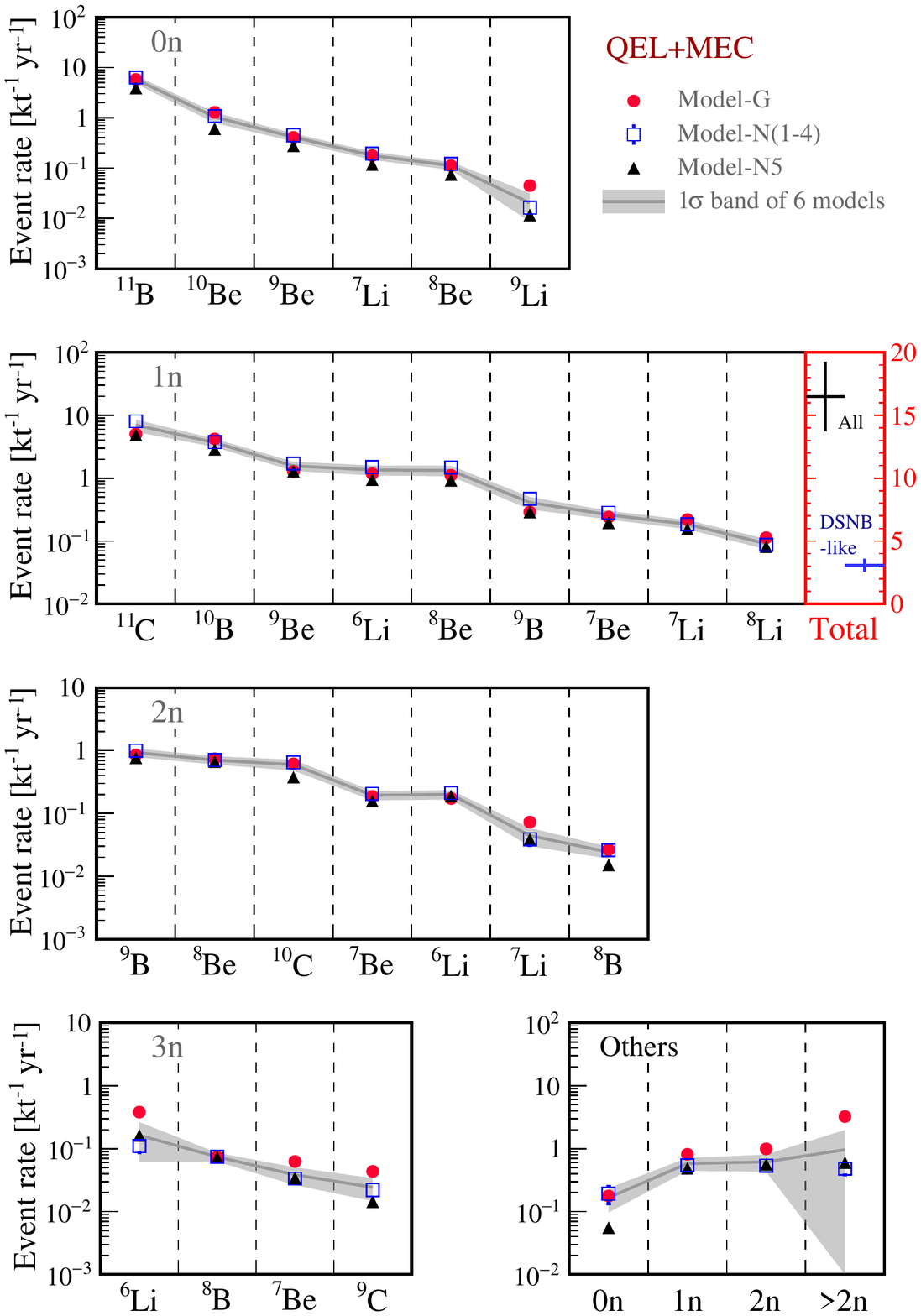}
\end{center}
\vspace{-0.6cm}
\caption{ The event rates for the NC interactions of atmospheric neutrinos (with energies ranging from 100 MeV to 10 GeV) with $^{12}$C nuclei in the exclusive channels, which are categorized by the associated neutron multiplicities. The predictions from Model-G, Model-N$i$ for $i = 1, 2, 3, 4$ and Model-N5 are denoted as dots, squares and triangles, respectively. Note that the squares with error bars stand for the mean value of the predictions from Model-N$i$ (for $i = 1, 2, 3, 4$) and $1\sigma$ deviation. In all the panels, the raw rates are extracted from those in Fig.~\ref{fig:table2}, and the gray bands represent the $1\sigma$ uncertainties from all six models. In the rightmost panel of the second row for a single neutron, the total rate $(16.5\pm 2.8)~{\rm kt}^{-1}~{\rm yr}^{-1}$ (black) and that $(3.1\pm 0.5)~{\rm kt}^{-1}~{\rm yr}^{-1}$ for the DSNB-like events (blue) should be read off from the vertical axis with red ticks.
\label{fig:dsnbrate2}}
\end{figure}

The $p$, $n$, $\alpha$, and $\gamma$-rays will deposit their kinetic energies in LS immediately after their production. If followed by a neutron capture, the prompt scintillation signals can mimic the prompt event of an IBD coincidence. The neutrons are tagged with high efficiency (better than $99\%$) via their captures on hydrogen in the LS detectors. Therefore, the event rates of the exclusive channels of the QEL process have been categorized in Fig.~\ref{fig:dsnbrate2} according to the neutron multiplicities. Some comments on Fig.~\ref{fig:dsnbrate2} are helpful.
\begin{itemize}
\item The event rates for all channels in the series of models (i.e., Model-N$i$ for $i=1, 2, 3, 4$) are quite similar, thus the average value and the standard deviation of the predictions from these four models are shown as squares with error bars, and labelled by ``Model-N(1-4)". However, the error bars are too small to be visible. Generally speaking, the event rate decreases as the neutron multiplicity increases. To visualize the model dependence, we have also shown the $1\sigma$ variation of all the six representative models as the gray shaded band.

\item Due to the high neutron tagging efficiency, NC interactions associated without neutrons or with more than one neutrons can be rejected as the background events in DSNB searches. Only single-neutron events have to be considered, for which the rates are given in the second row of Fig.~\ref{fig:dsnbrate2}. The rightmost panel of this row provides the total rate of all single-neutron interactions as $(16.5 \pm 2.8)~{\rm kt}^{-1}~{\rm yr}^{-1}$. When regarding only events in the prompt visible energy range from $11~{\rm MeV}$ to $30~{\rm MeV}$, most relevant for DSNB searches, the rate reduces to $(3.1 \pm 0.5)~{\rm  kt}^{-1}~{\rm yr}^{-1}$. Notice that the associated uncertainty is about $20\%$, representing the model variations of neutrino interactions. If the extra uncertainty of $15\%$ from the calculations of atmospheric neutrino fluxes is simply added in quadrature, one may obtain the overall uncertainty of $25\%$ for the NC backgrounds.

\item As indicated in the fourth row of Fig.~\ref{fig:dsnbrate2}, the \texttt{GENIE} generator produces significantly higher rate of the channels with more than two neutrons. In future large LS detectors, the neutron multiplicity distribution can be measured and {used to validate} the nuclear models. In addition, the decays of unstable final-state nuclei may provide unique signatures that allow for {\it in situ} measurements of the NC backgrounds. In a companion paper, we shall carry out a systematic study of such {\it in situ} measurements in the LS detectors~\cite{insitu}.
\end{itemize}

It is worthwhile to emphasize that the conversion of the kinetic energies of final-state particles in the NC interactions to the visible energies of the final events in the LS detectors is nontrivial. Such a conversion is process-dependent and relies on the energy deposition of different types of final-state particles. In our calculations of possible NC backgrounds for DSNB signals, we have employed the method of Monte Carlo simulations based on the software \texttt{GEANT4} (4.9.4)~\cite{Agostinelli:2002hh}, without including a specific detector geometry and the optical processes. The quenching effect of LS is considered according to the description in the Appendix of Ref.~\cite{An:2015jdp}. After converting the event rates in terms of kinetic energies of final-state particles into those of the quenched visible energies, we obtain the total rate of the IBD-like signals of the atmospheric neutrino NC interactions in the energy range from 11 to 30 MeV. Finally, one should notice that the typical {predicted} rate of DSNB signals in LS detectors is around $0.3~{\rm kt}^{-1}~{\rm yr}^{-1}$~\cite{An:2015jdp}, which is about one order of magnitude smaller than the atmospheric NC background rate $(3.1\pm 0.5)~{\rm kt}^{-1}~{\rm yr}^{-1}$. However, pulse shape discrimination offers a technique of efficient background suppression for non-positron prompt events, which will be necessary to ultimately achieve an unambiguous discovery of the DSNB signals~\cite{An:2015jdp}.

\subsection{Proton Decay}\label{subsec:pd}

Then, we turn to the detection of proton decay in the LS detectors, for which the signals and relevant backgrounds are quite different from those for DSNB. For the wCh detectors, the decay channel $p \to e^+ + \pi^0$ offers the clearest signature for proton decay~\cite{Miura:2016krn, Tanaka:2020emn}. However, as shown in Refs.~\cite{Sakai:1981pk, Weinberg:1981wj, Ellis:1981tv}, the supersymmetric minimal SU(5) GUTs may predict a highly suppressed decay rate for $p \to e^+ + \pi^0$, but an appreciably large one for $p \to K^+ + \overline{\nu}$, where the flavors of $\overline{\nu}$ depend on specific models~\cite{Murayama:2001ur} and are irrelevant for our following discussions. With a null signal in this channel, the SK experiment has placed a lower limit on the partial proton lifetime~\cite{Hayato:1999az, Kobayashi:2005pe} and the latest result is $\tau^{}_p/{\cal B}(p\to K^+\overline{\nu}) > 5.9\times 10^{33}~{\rm yr}$ at the $90\%$ confidence level~\cite{Abe:2014mwa}, where $\tau^{}_p$ is the proton lifetime and ${\cal B}(p\to K^+\overline{\nu})$ is the branching ratio.

As for proton decay in the channel $p \to e^+ + \pi^0$ in the LS detectors, the immediate scintillation light from $e^+$'s and the instantaneous decay of $\pi^0 \to 2\gamma$ lead to a single prompt signal, which can easily be contaminated by numerous backgrounds. In contrast, the decay channel $p \to K^+ + \overline{\nu}$ serves as the most sensitive probe for proton decay in the LS detectors. The main features of this decay channel and possible backgrounds are summarized as follows.
\begin{enumerate}
\item The $K^+$ meson decays quickly (with a lifetime $\tau^{}_{K^+} = 12.4~{\rm ns}$) into six channels, namely, $\mu^+ \nu^{}_\mu$ ($63.56\%$), $\pi^+ \pi^0$ ($20.67\%$), $\pi^+\pi^+\pi^-$ ($5.58\%$), $\pi^0 e^+ \nu^{}_e$ ($5.07\%$), $\pi^{0}\mu^{+}\nu_{\mu}$($3.35\%$) and  $\pi^+\pi^0\pi^0$ ($1.76\%$), where the corresponding branching ratios are given in the parentheses~\cite{Tanabashi:2018oca}. In order to identify the $K^+$ signal, one has to analyze these decay products and their signals in the LS detectors. The most important decay modes in LS are $K^+ \to \mu^+ \nu^{}_\mu$ and $K^+ \to \pi^+ \pi^0$, because they produce a signal of three-fold coincidence. See, e.g., Refs.~\cite{An:2015jdp},~\cite{Undagoitia:2005uu} and \cite{TheKamLAND-Zen:2015eva}, for earlier discussions. As $K^+$ is a heavy and highly ionizing charged particle in LS, it will {lose} its kinetic energy rapidly, producing a first prompt scintillation signal. In either decay mode there is a shortly delayed signal ($\sim\tau^{}_{K^+}$) from the daughter particle(s). In the first decay mode, the kinetic energy of $\mu^+$ constitutes the shortly delayed signal. Then the final-state $\mu^+$ decays into $e^+ \nu^{}_e \overline{\nu}^{}_\mu$ with a proper lifetime of $\tau^{}_{\mu^+} = 2.2~\mu{\rm s}$, which is long enough to be separated from the preceding two signals, and the Michel electron from the $\mu^+$ decay can be reconstructed as the third delayed signal. In the second decay mode, the neutral pion $\pi^0$ instantaneously decays into two gamma rays, while the charged pion $\pi^+$ decays primarily into $\mu^+ \nu^{}_\mu$ (with a proper lifetime of $\tau^{}_{\pi^+} = 26~{\rm ns}$).
The deposition of the total energy of the $\pi^0$ decay and the kinetic energy of $\pi^+$ are indistinguishable from each other, and will constitute the shortly delayed signal of this mode. Moreover, the daughter $\mu^+$ has low kinetic energy ($\sim$4 MeV) and its signal will be submerged in the tail of the shortly delayed signal. As a consequence, similar to the first decay mode, the signature of the second decay mode also represents a three-fold coincidence of the prompt signal, a shortly delayed signal and a single Michel electron. In the following discussions, we shall take the proton decay $p \to K^+ + \overline{\nu}$ and the subsequent decays $K^+ \to \mu^+ \nu^{}_\mu$ or $\pi^+ \pi^0$ as the target signal.
\begin{figure}[!t]
\begin{center}
\begin{tabular}{l}
\hspace{-0.0cm}
\includegraphics[width=0.75\textwidth]{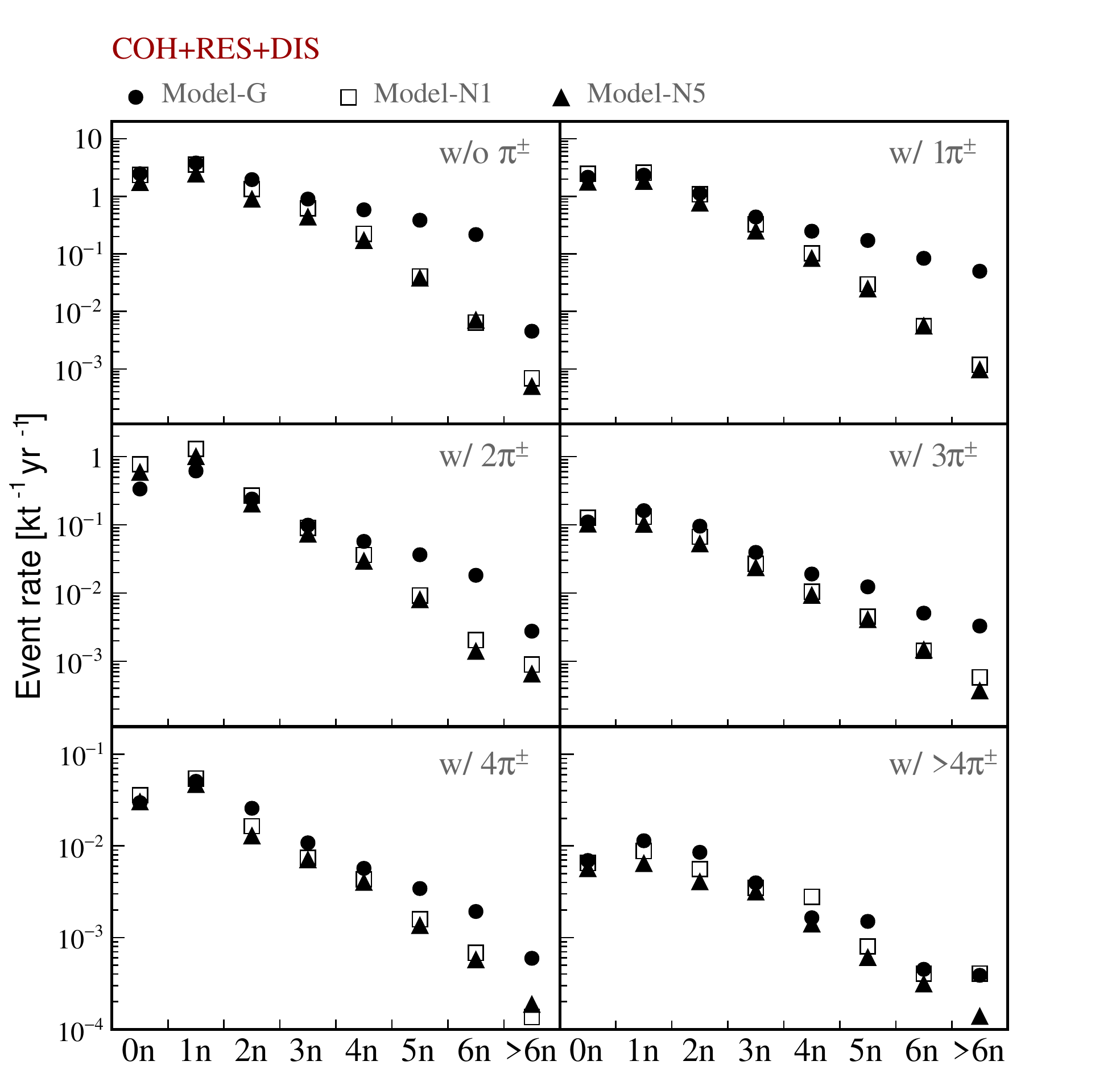}
\end{tabular}
\end{center}
\vspace{-0.6cm}
\caption{Neutron multiplicity distributions of the NC interactions of atmospheric neutrinos (with energies ranging from 100 MeV to 10 GeV) with $^{12}$C nuclei have been categorized by the multiplicities of the charged pions in the COH, RES, and DIS processes. The predictions simulated by using Model-G, Model-N1 and Model-N5 are denoted by dots, squares and triangles, respectively.}
\label{fig:neuPiMulti}
\end{figure}
\item Based on the above discussions about the signal of $p \to K^+ + \overline{\nu}$, we now discuss the possible background in the NC interactions of atmospheric neutrinos with $^{12}{\rm C}$ in the LS detectors. First, for the $K^\pm$ production, there will be a three-fold coincidence signal from the $K^\pm$ decay, but the associated production of other particles also contributes to the first prompt signal, which tends to have a relatively higher prompt energy compared to that of the three-fold signature of proton decay.
    Second, if a single neutral pion $\pi^0$ is produced, there will be no three-fold coincidence signals because of the prompt decay of $\pi^0$ into two $\gamma$'s.
    Third, for the $\pi^\pm$ production, the first prompt signal comes from the scintillation light of the $\pi^\pm$ kinetic energy and other associated particles,
    and the third delayed signal is the possible Michel electron from the muon decay. Therefore, to mimic the proton decay signal, the key is to have a suitable shortly delayed signal, which can be produced from the hadronic interactions of high energy $\pi^\pm$ and nucleons in LS if they are separable from the prompt scintillation signal. The possibility to achieve the separation heavily depends on the detailed simulation of particle interactions in LS, which will not be further explored in the current paper.
    Instead, we shall calculate the total rate of single $\pi^\pm$ production as a benchmark number.
    Finally, if one or more neutrons are produced in the NC interactions, high efficiency rejection power can be achieved using the typical neutron capture signature. Therefore we shall consider the possible production of single $\pi^\pm$ or $K^\pm$ but without neutrons as our focus in the following part.
    Note that a single negatively-charged pion $\pi^-$ will decay mainly into $\mu^- \overline{\nu}^{}_\mu$. Unlike $\mu^+$ which only decays, the stopped $\mu^-$ in LS will be captured into an atomic orbital and then it can either decay or undergo nuclear capture. The muon capture on $^{12}$C in LS may emit one or more neutrons. A recent measurement of muon capture on light nuclear isotopes in LS can be found in Ref.~\cite{Abe:2015wwn}.
%
\item In Fig.~\ref{fig:neuPiMulti}, neutron multiplicity distributions of the NC interactions of atmospheric neutrinos (with energies ranging from 100 MeV to 10 GeV) with $^{12}{\rm C}$ nuclei have been categorized by the multiplicities of the charged pions in the COH, RES, and DIS processes. Note that these multiplicity distributions can be measured in future large LS detectors and will be very useful to scrutinize the nuclear models. For proton decay, the visible energies of final-state particles will be as high as several hundred MeV, so the corresponding neutrino energies will be much higher, for which the COH, RES and DIS processes will be relevant. However, the COH process produces only $\pi^0$ that will not contribute to the background. Moreover, the three-fold coincidence signature consists of only one Michel electron from muon decay, so the processes associated with multiple $\pi^\pm$'s or neutrons are distinguishable from proton decay. For this reason, those with no neutron and only one charged pion are able to contaminate the proton decay signal. Note that $(1-2)\%$ of the $1\pi^{\pm}0n$ processes are accompanied by a charged kaon, but they can be rejected by the criterion of only one Michel electron. As a consequence, the rates for the potential background $0K^{\pm}1\pi^{\pm}0n$~\footnote{Here $0K^{\pm}1\pi^{\pm}0n$ denotes the channel that includes one charged pion, and meanwhile without $K^{\pm}$ and neutrons. There is no requirement on the presence or absence of other particles. It is similar for the notation of $1K^\pm 0\pi^\pm 0 n$.}
    from Model-G, Model-N1 and Model-N5 are found to be $2.1~{\rm kt}^{-1}~{\rm yr}^{-1}$, $2.4~{\rm kt}^{-1}~{\rm yr}^{-1}$ and $1.8~{\rm kt}^{-1}~{\rm yr}^{-1}$, respectively,  where the average and 1$\sigma$ deviation of these models is $(2.10\pm 0.27)~{\rm kt}^{-1}~{\rm yr}^{-1}$.
\begin{figure}[!t]
\begin{center}
\begin{tabular}{l}
\hspace{-0.0cm}
\includegraphics[width=0.6\textwidth]{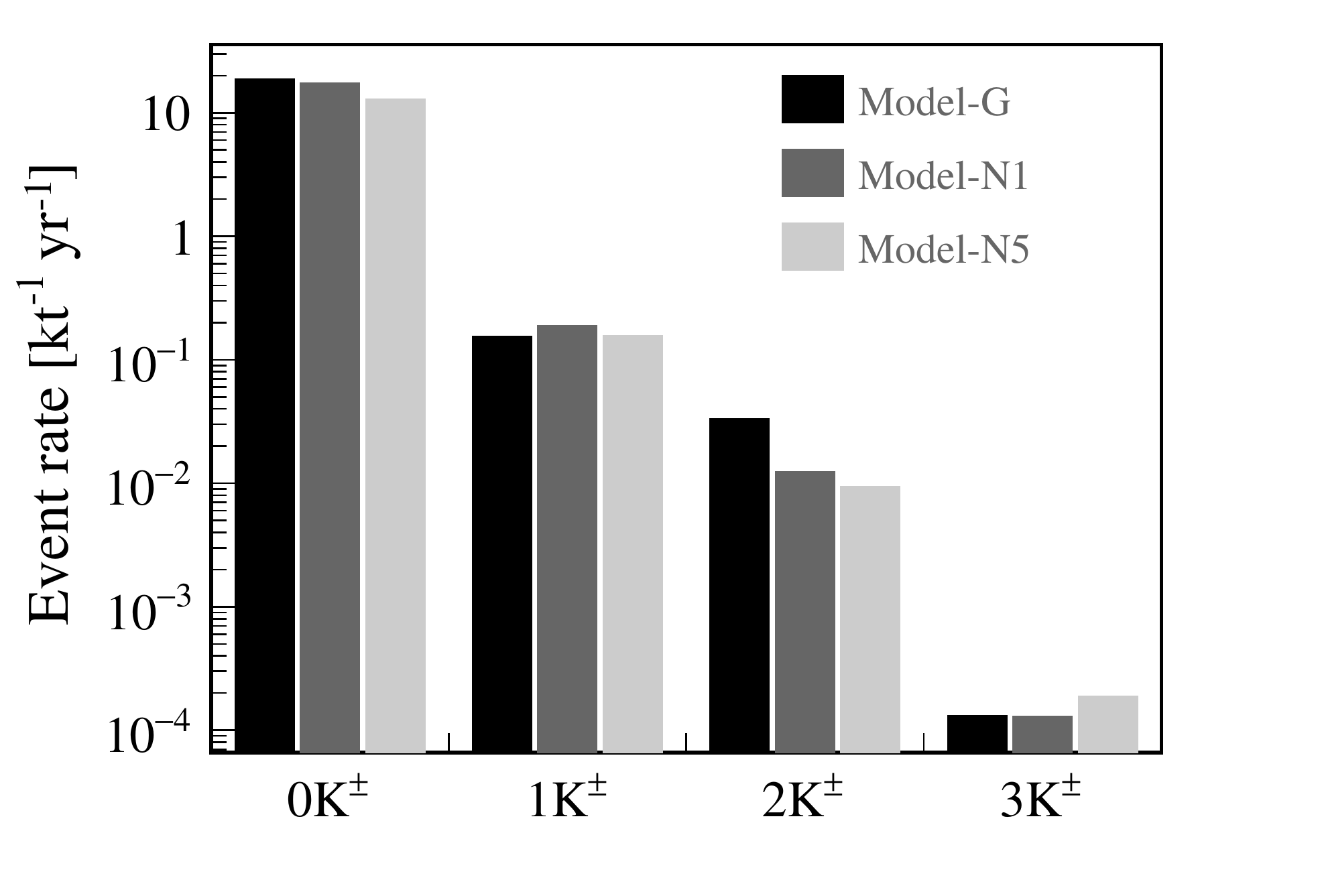}
\end{tabular}
\end{center}
\vspace{-0.6cm}
\caption{The multiplicity distribution of charged kaons from the NC interactions of atmospheric neutrinos (with energies ranging from 100 MeV to 10 GeV) with $^{12}{\rm C}$ nuclei in the COH, RES, and DIS processes. Note that there are no NC events with more than three charged kaons in Model-G, Model-N1 and Model-N5.}
\label{fig:kaontot}
\end{figure}
\item In Fig.~\ref{fig:kaontot}, the multiplicity distribution of the produced charged kaons from the COH, RES, and DIS processes is shown. To illustrate the model dependence, we have performed the calculations by using Model-G, Model-N1 and Model-N5, the results for which are represented by {black, dark-grey and grey histograms}, respectively. {Note that the calculations of Model-N$i$ for ($i = 1, 2, 3, 4$) are practically identical in the COH, RES, and DIS processes.} The average fractions of NC events from these processes without charged kaons and with a single charged kaon are about $98.9\%$ and $1.0\%$, respectively. The low rate of kaon production is due to the fact that a relatively high energy for the incident neutrino is required and the event rate above a neutrino energy of 1 GeV decreases rapidly, as shown in Fig.~\ref{fig:eventrate}. The event rate for a single $K^\pm$ production (i.e., $1K^\pm 0\pi^\pm 0 n$) can be estimated as $0.057~{\rm kt}^{-1}~{\rm yr}^{-1}$, $3.8\times 10^{-3}~{\rm kt}^{-1}~{\rm yr}^{-1}$ and $3.1 \times 10^{-3}~{\rm kt}^{-1}~{\rm yr}^{-1}$ from Model-G, Model-N1 and Model-N5, respectively. The average and standard deviation of these three models is then $(0.021^{+0.025}_{-0.021})~{\rm kt}^{-1}~{\rm yr}^{-1}$, where the uncertainty comes from the large variation of model predictions between \texttt{GENIE} and \texttt{NuWro} (see the right panel of Fig.~\ref{fig:kaon}).

\begin{figure}
\begin{center}
\begin{tabular}{l}
\hspace{-0.5cm}
\includegraphics[width=0.55\textwidth]{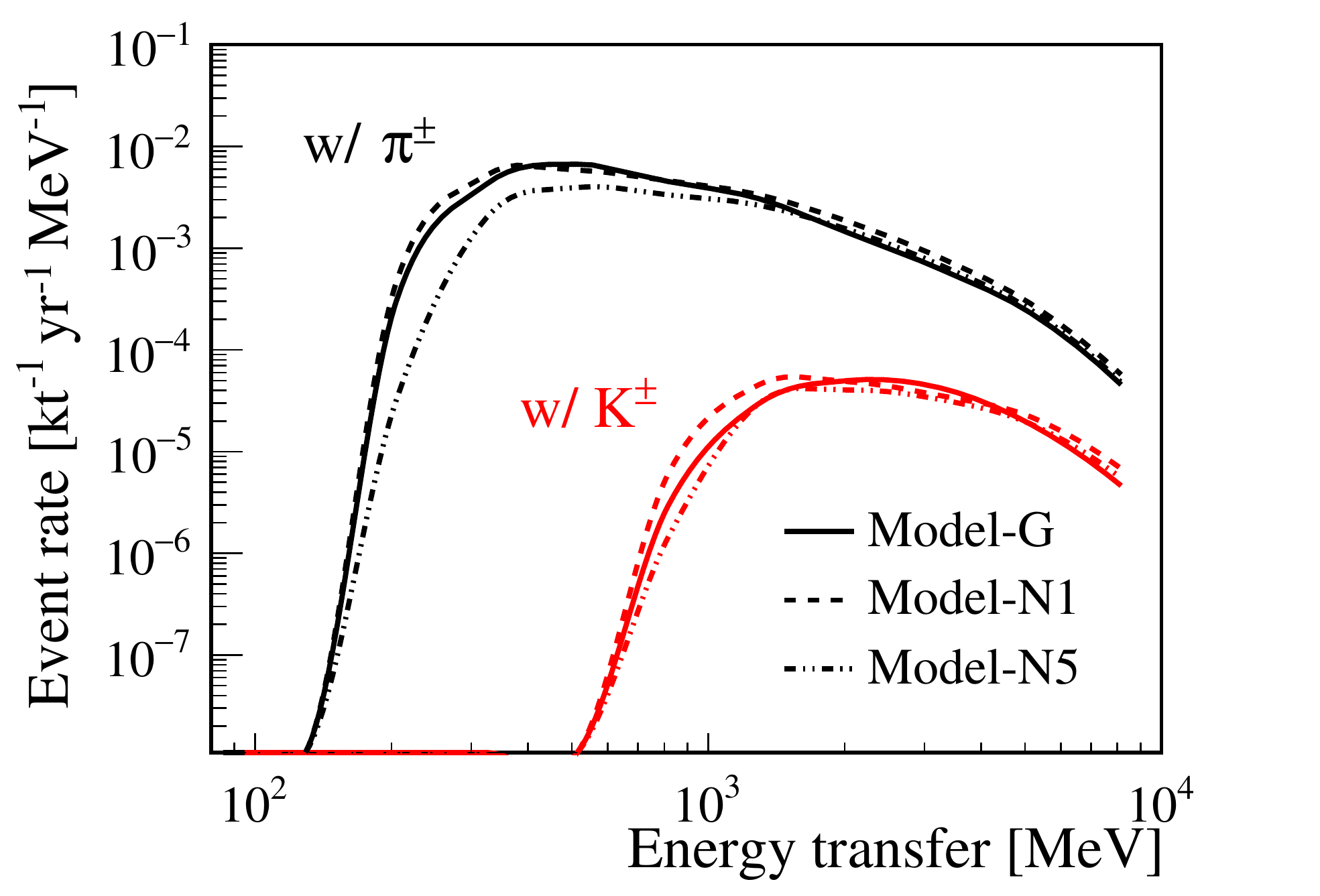}
\hspace{-0.9cm}
\includegraphics[width=0.55\textwidth]{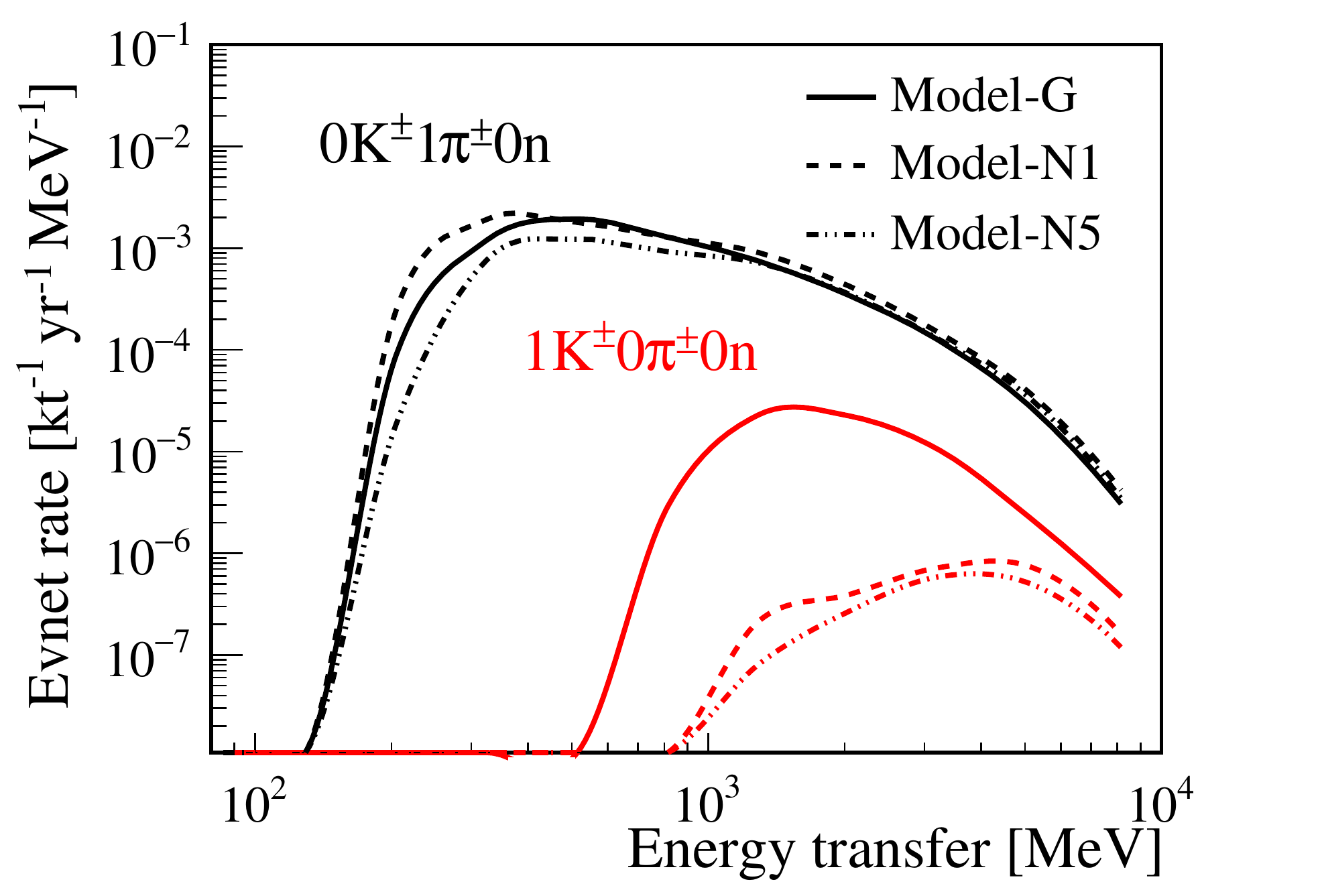}
\end{tabular}
\end{center}
\vspace{-0.6cm}
\caption{The differential event rate for the production of $\pi^\pm$ and $K^\pm$ with respect to the energy transfer, where the results of any number of $\pi^\pm$ and $K^\pm$ are given in the left panel and those of the exclusive one single $\pi^\pm$ or $K^\pm$ are shown in the right panel. Note that only Model-G and Model-N1 have been implemented to illustrate the difference between \texttt{GENIE} and \texttt{NuWro} generators.}
\label{fig:kaon}
\end{figure}
\item The differential rates of the interactions associated with the production of $\pi^\pm$ and $K^\pm$ are shown in Fig.~\ref{fig:kaon} with respect to the energy transfer, which is defined as the energy difference between incoming and outgoing neutrinos. {The results of total $\pi^\pm$ and $K^\pm$ production and the exclusive processes $0K^{\pm}1\pi^{\pm}0n$ and $1K^{\pm}0\pi^{\pm}0n$ are illustrated in the left and right panels respectively.
    Some important observations can be made. First, the $\pi^\pm$ production starts with a relatively low energy transfer of around 150 MeV, whereas the $K^\pm$ production needs much higher energy transfers of above 500 MeV. The differential rates of total $\pi^\pm$ and $K^\pm$ production are compatible between GENIE and NuWro, in spite of the different classifications of the RES and DIS processes in event generators. Second, for the exclusive $0K^{\pm}1\pi^{\pm}0n$ and $1K^{\pm}0\pi^{\pm}0n$ production, the differential event rates are similar for the former one, but rather distinct for the $1K^{\pm}0\pi^{\pm}0n$ process, where GENIE predicts much higher production rate than all versions of NuWro. {For the $K^\pm$ production in NuWro, there are higher possibilities to be associated with the appearance of $\pi^\pm$ or neutrons, which lead to a much lower branching ratio for the $1K^{\pm}0\pi^{\pm}0n$ process.}
     Third, comparing between the left and right panels, we can see that the single $\pi^\pm$ process dominates over the multiple $\pi^\pm$ in both \texttt{GENIE} and \texttt{NuWro}. Nevertheless, while the single $K^\pm$ production without neutrons and $\pi^\pm$'s is most important in the total $K^\pm$ production of \texttt{GENIE}, it becomes unimportant in \texttt{NuWro}. Such differences might arise from different treatments of the transition from RES to shallow and deep inelastic scattering processes, and surely deserves further careful investigation.
}
\end{enumerate}

As indicated in Fig.~\ref{fig:kaon}, the energy transfer for kaon production is at least $500~{\rm MeV}$, so it is possible to apply an energy cut to efficiently reduce the background associated with kaons. {In a realistic LS detector, apart from the energies carried away by invisible final-state particles, all the kinetic energies of visible particles would be deposited, but the conversion of the energy transfer to the visible energy is nontrivial and heavily depends on the types of final-state particles.} In this paper we directly apply the energy cut on the energy transfer to obtain a rough estimation, and leave the detailed detector response modeling for future works by the experiments. We take the energy transfer range from $150~{\rm MeV}$ to $650~{\rm MeV}$ for the single $\pi^\pm$ production ($0K^{\pm}1\pi^{\pm}0n$),
and the rate is reduced to $(0.61\pm 0.14)~{\rm kt}^{-1}~{\rm yr}^{-1}$. On the other hand, since a large proportion of the rest mass of $K^\pm$ will be carried away by invisible neutrinos from kaon decay, we require the energy transfer to be ranging from $150~{\rm MeV}$ to $1~{\rm GeV}$ for the single $K^\pm$ production ($1K^{\pm}0\pi^{\pm}0n$). Then the corresponding rate will be $(4^{+5}_{-4}\times 10^{-4})~{\rm kt}^{-1}~{\rm yr}^{-1}$.

Here we would like to remind that not all the single $\pi^\pm$ events, but only those being recognized as the three-fold coincidence signal will constitute the background of proton decay. In this aspect, we need careful treatments on the particle interaction in LS and detector simulation. The pulse shape discrimination of the single $\pi^\pm$ events and even advanced deep learning techniques~\cite{Abratenko:2020pbp} would be critical to obtain a reasonable background level for proton decay. Such further exploration will be reported in future works.
Moreover, it should be noted that energetic neutrons and protons induced by neutrino interactions with $^{12}$C may produce secondary $\pi^{\pm}$'s. Such interactions can also contribute to the backgrounds for searches of proton decay. Additionally, we focus on the NC interactions of atmospheric neutrinos and leave out the CC interactions, which should also be taken into account in more realistic background analysis. It is clear that our calculation strategy of the NC background will play an important role in such an analysis as well.


\section{Summary}\label{sec:summary}

In this paper, we have performed a systematic calculation of the NC background induced by atmospheric neutrino interactions with $^{12}{\rm C}$ nuclei in the LS detectors, which are expected to be crucially important for the experimental searches for DSNB and proton decay. In our calculations, the up-to-date fluxes of atmospheric neutrinos at the JUNO site provided by the Honda group are used. As for neutrino-nucleus interactions, we have chosen six {typical} models from the Monte Carlo neutrino event generators \texttt{GENIE} and \texttt{NuWro} and take the model variation as the systematic uncertainty in the background prediction. Then, a statistical configuration model of $^{12}{\rm C}$ is implemented to determine the probability distribution of the excited states of final-state nuclei produced in the NC interactions. The deexcitation processes of these nuclei are handled by \texttt{TALYS}. Taking account of neutrino interactions and deexcitation processes, we are able to compute event rates of exclusive processes with final-state protons, neutrons, $\gamma$'s and $\alpha$'s and the corresponding energy spectra. These processes are relevant for the experimental searches for the IBD signals of DSNB $\overline{\nu}^{}_e$'s. In addition, the production of multiple neutrons, kaons and pions in the high-energy region is considered, and {the} implications for the detection of proton decay are investigated.

For the detection of DSNB in LS detectors, the golden channel is the IBD process. {To reject} the irreducible backgrounds from reactor $\overline{\nu}^{}_e$'s in the low-energy region and atmospheric $\overline{\nu}^{}_e$'s in the high-energy region, one has to focus on the narrow window of the visible energies $11~{\rm MeV} \lesssim E^{}_{\rm vis} \lesssim 30~{\rm MeV}$. In this energy window, the NC backgrounds induced by atmospheric neutrinos will be dominant. As the visible energy is relatively low, the QEL process of neutrino-$^{12}{\rm C}$ interactions turns out to be most important. Though the neutron tagging efficiency is intrinsically high for LS detectors, {the single-neutron NC backgrounds} lead to IBD-like signals and thus are irreducible. {Using the numerical simulations from two neutrino event generators} (i.e., \texttt{GENIE} and \texttt{NuWro}) and \texttt{TALYS} for deexcitation, we have found the event rate for the exclusive processes with one neutron production to be $(16.5\pm 2.8)~{\rm kt}^{-1}~{\rm yr}^{-1}$ in the whole range of visible energies, where the uncertainty originates from different models for neutrino-$^{12}{\rm C}$ interactions. When restricted into the energy window $11~{\rm MeV} \lesssim E^{}_{\rm vis} \lesssim 30~{\rm MeV}$ of interest, the event rate is $(3.1\pm 0.5)~{\rm kt}^{-1}~{\rm yr}^{-1}$. Further reduction of the NC backgrounds from atmospheric neutrinos can be achieved by using pulse shape discrimination~\cite{An:2015jdp}.

For the proton decay in LS detectors, the decay mode $p \to K^+ + \overline{\nu}$ is most promising. The three-fold coincidence {from} the energy deposition of $K^+$, the shortly delayed signal of $\mu^+$ from $K^+ \to \mu^+ \nu^{}_\mu$ (or $\pi^+\pi^0$ with $\pi^+ \to \mu^+\nu^{}_\mu$), and the single Michel electron from $\mu^+ \to e^+\nu^{}_e \overline{\nu}^{}_\mu$, {will constitute a unique signature for this decay mode.} We identify the relevant backgrounds from NC interactions of atmospheric neutrinos as the production of {a single charged pion or kaon}.  When requiring the energy transfer to be ranging from $150~{\rm MeV}$ to $650~{\rm MeV}$ for the single pion production (i.e., $0K^\pm 1\pi^\pm 0n$) and from $150~{\rm MeV}$ to $1~{\rm GeV}$ for the single kaon production (i.e., $1K^\pm 0\pi^\pm 0n$), we find that the background rates are
$(0.61\,\pm\,0.14)~{\rm kt}^{-1}~{\rm yr}^{-1}$ and $(4^{+5}_{-4}\times 10^{-4})~{\rm kt}^{-1}~{\rm yr}^{-1}$ respectively,
where the uncertainty arises from the model variation of neutrino interactions. Therefore, further examination of all possible backgrounds for proton decay in the LS detectors must pay a particular attention to the NC events from atmospheric neutrinos, as we have demonstrated.

Apart from the future ordinary LS detectors~\cite{An:2015jdp, Wurm:2011zn}, we will have other large-scale detectors with advanced techniques based on water~\cite{Abe:2018uyc}, water-based LS~\cite{Askins:2019oqj}, {and} liquid-Argon~\cite{Bueno:2007um, Acciarri:2015uup}. The experimental searches for DSNB and proton decay are also primary physics goals for these detectors, and the NC backgrounds induced by atmospheric neutrinos will be relevant. We believe that the calculations performed in the present work will be not only useful for the LS detectors, but also instructive for the parallel studies for other types of detectors.

\section*{Acknowledgements}
The authors are greatly indebted to Morihiro Honda for his help in the calculation of atmospheric neutrino fluxes, and to Dong-liang Fang, Yoshinari Hayato and Julia Sawatzwi for their helpful discussions. We also would like to thank Wan-lei Guo and Michael Wurm for carefully reading the manuscript and valuable comments. This work was supported in part by the National Key R$\&$D Program of China under Grant No.~2018YFA0404100, by National Natural Science Foundation of China under Grant No.~11835013, by the Strategic Priority Research Program of the Chinese Academy of Sciences under Grant No.~XDA10010100, by the CAS Center for Excellence in Particle Physics (CCEPP), and by the China Postdoctoral Science Foundation funded project under Grant No. ~2019M660793.


\end{document}